\documentclass[aps,prd,10pt,nofootinbib,twocolumn,eqsecnum,showpacs,showkeys,superscriptaddress,preprintnumbers,altaffilletter,floatfix]{revtex4-2}
\usepackage{graphicx}
\usepackage[T1]{fontenc}
\usepackage[utf8]{inputenc}
\usepackage{amsmath}
\usepackage{multirow}
\usepackage{hyperref}
\usepackage{soul}
\usepackage{graphicx}
\usepackage{dcolumn}
\usepackage{amssymb}
\usepackage{amsfonts}
\usepackage{amsbsy}
\usepackage{color}
\usepackage{xcolor}
\usepackage{rotating}
\usepackage[english]{babel}
\usepackage{multirow}
\usepackage{float}
\usepackage{placeins}  

\newcommand{\be}{\begin{equation}}
\newcommand{\ee}{\end{equation}}
\newcommand{\bea}{\begin{eqnarray}}
\newcommand{\eea}{\end{eqnarray}}
\usepackage{booktabs,makecell}
\newcolumntype{Y}{>{\centering\arraybackslash}X}

\usepackage{enumitem} 
\usepackage{booktabs}        
\usepackage{tabularx}        
\usepackage{siunitx}         
\newcolumntype{Y}{>{\centering\arraybackslash}X}  

\hypersetup{
colorlinks=true,
linkcolor=blue,
filecolor=magenta,
urlcolor=cyan,
citecolor=cyan
}

\begin{document}

\title{Growth of Cosmic Structures in generalized mass-to-horizon relation Entropic Cosmology}
\author{Shafqat Ali}
\email{shafqat.ali@phd.usz.edu.pl}
\affiliation{Institute of Physics, University of Szczecin, Wielkopolska 15, 70-451 Szczecin, Poland}
\author{T. Denkiewicz}
\email{tomasz.denkiewicz@usz.edu.pl}
\affiliation{Institute of Physics, University of Szczecin, Wielkopolska 15, 70-451 Szczecin, Poland}

\date{\today}

\begin{abstract}
We investigate the growth of cosmic structures in the thermodynamically consistent \emph{generalised mass-to-horizon} entropic cosmology (MHEC).  For the Bekenstein case $n=1$ the entropic energy density $\rho_{e}=\gamma H^{2}/(4\pi G)$ augments the Friedmann equations without modifying the Hawking temperature and automatically satisfies the Clausius relation, thereby avoiding the inconsistencies that afflicted earlier entropy–force models.  We then derive the linear perturbation equations, emphasising the distinction between a \emph{fully perturbed} interaction term (Case~A) and the common approximation in which the perturbation is neglected (Case~B).  Numerical solutions show that Case~A follows the $\Lambda$CDM matter–growth history within the current $f\sigma_{8}$ uncertainties for $\gamma\lesssim10^{-2}$.  Our results demonstrate that MHEC matches both background and growth probes as well as $\Lambda$CDM without extra free parameters, providing a viable entropic explanation for recent accelerated expansion of the Universe.
\end{abstract}

\maketitle
 \section*{Introduction}
\noindent

The accelerated expansion of the Universe~\cite{Riess1998,Perlmutter1998} remains one of the most profound puzzles in modern cosmology.  
While a cosmological constant \(\Lambda\) offers a simple phenomenological fit to current observations, it is plagued by the “fine–tuning’’ and “coincidence’’ problems~\cite{Weinberg1989,Carroll2001}.

To mitigate these difficulties one may either modify gravity itself—through, for instance, \(f(R)\) or related extensions~\cite{NOJIRI2007,Nojiri2003,Capozziello2011,Hu2007,Starobinsky2007,Appleby2007}—or promote the dark‐energy sector to a genuinely dynamical component.  
In the latter case the accelerated expansion can be driven by scalar fields, giving rise to quintessence‐ or phantom‐like behaviour~\cite{Caldwell,Caldwell2002}, or by scalar–tensor theories in which gravity is mediated by both a tensor and a scalar degree of freedom~\cite{Brans1961}, some of which emerge as low‐energy limits of string theory.

An alternative line of inquiry links dark energy to horizon‐scale physics and quantum‐gravity principles.  
A prominent example is provided by the \emph{holographic principle} (HP)~\cite{Hooft1993,Susskind1995}, according to which all information contained in a spatial volume can be encoded on its boundary.  
\emph{Holographic dark‐energy} (HDE) models~\cite{Li2004,Wang2017} exploit this idea by tying the dark‐energy density to an infrared cutoff defined by a cosmological horizon, typically adopting the Bekenstein–Hawking entropy.   
Although HDE connects dark energy with horizon thermodynamics, it should not be confused with \emph{entropy‐based} or \emph{entropic} dark‐energy models, which start instead from \emph{modified} entropy–area laws and may therefore violate the first law of thermodynamics unless further consistency conditions are imposed.

Among the wide spectrum of proposals there are also dynamical vacuum models motivated by quantum-field theory in curved space-time~\cite{Sola2008,Shapiro2009,Sola2011}.  
Remarkably, these scenarios reproduce the same effective Friedmann equations that arise—by a completely different route—in entropic-cosmology frameworks, where an entropic energy density is induced by non-standard entropy–area relations on the Hubble (or apparent) horizon~\cite{Easson2011,EASSON2012}.  

It is important, however, to distinguish entropic cosmology from Verlinde’s entropic-gravity programme.
In entropic cosmology one keeps General Relativity intact and merely supplements the Einstein equations with extra, horizon-sourced “entropic-force” terms; gravity itself remains a fundamental interaction, and the usual field-theoretic perturbation machinery applies.
Verlinde’s approach, by contrast, re-interprets the entire gravitational interaction as an emergent entropic force derived from holographic information on screens, so that the Einstein equations appear only as large-scale, thermodynamic equations of state rather than fundamental field equations.
Consequently, observational signatures, the role of perturbations and even the conceptual status of geometric quantities differ in the two pictures \cite{Gohar2023}.
Representative realizations of the entropic-cosmology idea employ Tsallis-type horizon entropies \cite{Tsallis1988}.
Such models modify the Hubble-scale dynamics in a way that can ease persistent tensions in late-time cosmology, notably the \(H_0\) and \(S_8\) discrepancies, while still reducing to $\Lambda$CDM for suitable parameter choices.

However, growth–of–structure tests place severe pressure on the \emph{original} entropic–force proposals in which cosmic acceleration is driven solely by terms proportional to \(H\), \(\dot H\) or \(H^{n}\) without an additive vacuum component.  
Detailed comparisons with redshift–space–distortion and cluster data show that such models systematically under–predict the observed growth rate and are now disfavoured at more than \(3\sigma\) confidence~\cite{Koivisto2011,Basilakos2012,Basilakos2014}.  

In this work we revisit entropic cosmology within the \emph{thermodynamically consistent} generalized mass-to-horizon relation framework.  
Besides reproducing the exact \(\Lambda\)CDM background when \(n=3\), we find that the \textit{Bekenstein limit} \(n=1\) already delivers an excellent simultaneous fit to current data:  
(i) the dark–matter density contrast \(\delta_m(k,a)\) and the derived growth observable \(f\sigma_{8}(z)\) evolve virtually indistinguishably from their \(\Lambda\)CDM counterparts;  
(ii) the deceleration parameter \(q(z)\) and the fractional energy densities \(\Omega_i(z)\), $(i=m,r)$ follow the standard trajectories; and  
(iii) the predicted Hubble history \(H(z)\) matches cosmic–chronometer within the 1\(\sigma\) error bars and remains consistent with the BAO distance ratios used in Ref. \cite{Gohar2024}.  

The same generalized MHEC model has already passed stringent tests against purely \emph{geometrical} data—SNIa, BAO, cosmic chronometers and CMB distance priors—in Ref.~\cite{Gohar2024}.  Here we go one step further by, for the first time, confronting it with growth–of–structure observations, providing the \emph{dynamical} probe of entropic energy density.

The structure of the paper is as follows. In Section \ref{sec:themodel} we introduce the generalized mass-to-horizon relation entropic cosmology (MHEC) framework, derive the generalized entropy implied by the mass–to–horizon relation, and check its thermodynamic consistency.
In Section \ref{sec:perturbations} we set up the linear perturbation equations for matter and radiation, outline the two alternative prescriptions for the interaction source, and obtain the master growth system that will be solved numerically.
Section \ref{sec:discussion} confronts the model with current background and growth data.
Finally, Section \ref{sec:conclusions} summarizes our main results and points to future extensions.

\section{The model}\label{sec:themodel}
The recent model proposed by Gohar and Salzano – (MHEC) \cite{Gohar2024} represents a novel approach within entropic cosmology frameworks, significantly differing from traditional models through its explicit inclusion of interaction terms and a generalized dependence on the Hubble parameter. What distinguishes the  model is its inherently dynamic and scale-dependent interaction terms, which couple dark matter, radiation, and entropic energy uniquely, thus providing a richer phenomenological landscape compared to previous entropic cosmology models. Unlike earlier cosmological perturbation frameworks, which typically treat interactions between cosmic fluids as small corrections or ignore them entirely, the model explicitly incorporates scale-dependent interactions into the fundamental continuity equations.

However, in such approaches, thermodynamic consistency is often compromised unless the horizon temperature is also modified. To resolve this issue, Gohar and Salzano~\cite{Gohar2024} recently proposed a \emph{generalized mass–horizon relation},
\begin{equation}
  M = \gamma \frac{c^2}{G} L^n \,, \label{mhr}  
\end{equation}
where \(n\) is a real parameter and \(\gamma\) encodes dimension-dependent factors. This framework leads to a new entropy scaling \(S \propto L^{\,n+1}\) that recovers Bekenstein entropy for \(n=1\) and generates a wide family of entropy-driven dark-energy models without altering the Hawking temperature. Remarkably, for \(n=3\) the background dynamics reduce exactly to those of \(\Lambda\)CDM.  
In this work, we build upon the MHEC approach to explore the cosmological implications of a thermodynamically consistent, entropy-based model of entropic energy.

Using the Clausius relation along with the Hawking temperature, the mass-to-horizon (\ref{mhr}) relation implies a generalized entropy:
\begin{equation}\label{genentropy}
    S_n=\gamma\frac{2n}{n+1}L^{\,n-1}S_{BH}
\end{equation} 
where $S_{BH}$ is the standard Bekenstein--Hawking entropy.
For specific values of $n$ and $\gamma$, this relation can recover various known entropy forms:
\begin{itemize}
\item Bekenstein entropy for $n=1$,
\item Tsallis--Cirto entropy for certain conditions $(1+n=2\delta)$,
\item Barrow entropy for $n=1+\Delta$ ($1\le n\le 2$).
\end{itemize}

The case $n=3$ yields a cosmology exactly equivalent to the $\Lambda$CDM model, meaning the entropic term acts precisely as a cosmological constant, thus providing a thermodynamic explanation for dark energy.

Other values of $n$ allow exploring a wider range of cosmological behaviors, potentially offering explanations for different dark energy phenomena and allowing observational differentiation.

A crucial point is that only certain temperatures (specifically the Hawking temperature) have solid theoretical background from quantum field theory. Hence, consistency with the Hawking temperature, enabled by the generalized mass--horizon relation, provides a robust theoretical footing for applying generalized entropies in cosmology.

The generalized mass-to-horizon relation enriches the theoretical and phenomenological landscape of holographic entropic energy models:
\begin{itemize}
    \item Providing a thermodynamically consistent framework connecting entropy, temperature, and horizon mass,
    \item Offering a physical interpretation of dark energy terms as emerging directly from horizon thermodynamics,
    \item Allowing various entropy scenarios (Tsallis, Barrow, etc.) to coexist within a single consistent framework, 
    \item Reinforcing the thermodynamic origin of the cosmological constant and other ``dark energy'' phenomena through observational consistency checks.
\end{itemize}

This innovative approach offers profound implications for the evolution of cosmic structures, particularly concerning the growth rate of perturbations across different cosmological epochs. While the Newtonian approximation has been historically employed for simplicity in many perturbative analyses, the MHEC model presents a unique scenario where interactions explicitly appear within the perturbation equations, challenging conventional methods and inviting a deeper examination into the limitations and potentials of the Newtonian approximation. This distinctive aspect of the MHEC model motivates our detailed investigation into how accurately Newtonian perturbation theory can capture the evolution of cosmic structures when interactions are non-trivial and scale-dependent.

In this article we rigorously analyze the growth of dark-matter and radiation perturbations within the Newtonian framework introduced by Gohar and Salzano, systematically outlining its strengths and limitations. Results are encouraging: over the redshift range probed by current redshift-space-distortion and cluster measurements, the entropic-energy-density model can reproduce the observed growth-rate data at a level comparable to $\Lambda$CDM. This success motivates a comprehensive Markov-Chain Monte Carlo study that will confront the full cosmological data set, a task we defer to forthcoming work.

\subsection*{Thermodynamic Consistency of the MHEC}
\label{sec:consistency}

\noindent
Below we summarise why the generalised mass–to-horizon relation entropic energy scenario is internally consistent from a thermodynamic standpoint.

\begin{enumerate}
\item \textbf{Clausius relation}  
      Using the Hawking temperature $T_H=\hbar c/(2\pi k_B L)$, together with the generalised mass–horizon relation (\ref{mhr}) the Clausius identity $dE=c^{2}\,dM=T\,dS$
      integrates to the entropy (\ref{genentropy}) proving that the set \(\{T_H,S_{n}\}\) is a consistent \((T,S)\) pair.

\item \textbf{Legendre (scaling) structure}  
      With \(d=3\) spatial dimensions, the scaling exponents satisfy
      \(\epsilon=\theta+d\) where \(U\propto L^{\,n}\) (\(\epsilon=n\)) and
      \(T\propto L^{-1}\) (\(\theta=-1\)), so that \(\epsilon=\theta+d\Rightarrow n=2\).  
      For arbitrary \(n\) the Legendre transformation still closes because the conjugate variables are constructed from \(S_{n}\) itself, ensuring extensivity for the special case \textbf{\(n=2\)} and proper non-extensive behaviour otherwise.

\item \textbf{Mass–horizon compatibility}  
      The same power \(n\) appears in both \(M(L)\) and \(S_{n}(L)\); hence no hidden rescalings are needed.  In particular, the parameter \(\gamma\) always enters as a single global prefactor, so the equality \(dE=c^{2}\,dM\) holds identically for every \(n\).
\item \textbf{Entropy–force closure}  
      From \(T_{H}\) and \(S_{n}(L)\) one obtains  
      \[
        F \;=\; -\,T_{H}\,\frac{dS_{n}}{dL}
          \;=\; -\,\gamma\,n\,\frac{c^{4}}{G}\,L^{\,n-1}.
      \]
      The force therefore depends on the two model parameters and scales as \(F\!\propto\!L^{\,n-1}\); it reduces to a true constant \(\bigl(|F|=\gamma c^{4}/G\bigr)\) only in the Bekenstein limit \(n=1\).

\end{enumerate}
\medskip
Consequently,
\begin{eqnarray}
U&\propto& L^{\,n}\;(\epsilon=n), \\
T&\propto& L^{-1}\;(\theta=-1),\\
\epsilon&=&\theta+d\ \Longrightarrow\ n=2,
\end{eqnarray}
so that extensivity is ensured precisely for \(n=2\), fully in agreement with
the generalized--mass framework of \cite{Gohar2023} and  \cite{Gohar2024}.

Because conditions 1–4 are simultaneously fulfilled, the MHEC setup avoids the inconsistencies highlighted in earlier entropic-cosmology models that mixed non-extensive entropies with an unmodified Hawking temperature \cite{Da̧browski2015,Gohar2024, Gohar2023}.  

\subsection{Background diagnostics}\label{subsec:bg-plots}
Before we turn to the density–perturbation sector we first inspect the model’s \emph{purely geometric diagnostics}—the expansion rate \(H(z)\), deceleration parameter \(q(z)\), effective dark-energy equation of state \(w_{\mathrm{DE}}(a)\) and fractional energy densities \(\Omega_i\).
These are exactly the observables that were successfully fitted to SNIa, BAO and cosmic-chronometer data in the geometry-only analysis of Ref.\,\cite{Gohar2024}, which did not yet include growth information.

To introduce the entropic force terms from $S_n$, for Bekenstein entropy n = 1, one can easily derive the following expression for entropic force $F_{n=1}$ and entropic pressure $p_{n=1}$. From now, we will use $F_{n=1}=F_{e}$, $p_{n=1}=p_{e}$
\begin{equation}
F_e = -\gamma \frac{c^4}{G},\qquad ~p_e = -\gamma \frac{c^4}{4\pi G}L^{-2}. \label{eq:entropic_F_p}
\end{equation}
The generalized entropic pressure $p_e$ can be written as: 
\begin{equation}
p_e = -c^2 \rho_e, \label{eq:p_rho}
\end{equation}
and the entropic energy density definition for $n=1$, $\rho_{n=1}=\rho_e$ as  
\begin{equation}
\rho_e = \gamma \frac{H^2}{4\pi G}. \label{rho_n}
\end{equation}

In order to introduce the entropic contributions in the Friedmann, acceleration, and continuity equations, we have \cite{Komatsu2014}
\begin{align} 
&H^2 =\frac{8\pi G}{3}\sum_{i} \rho_i + f(t) \,, \label{eq:Fa11}\\
&\frac{\ddot a}{a}=-\frac{4\pi G}{3}\sum_{i}\left(\rho_i+\frac{3p_i}{c^2}\right) + g(t) \,, \label{eq:Aa11} \\
&\sum_{i}\left[\dot \rho_i+3H\left(\rho_i+\frac{p_i}{c^2}\right)\right] =\frac{3H}{4\pi G}\left(-f(t)-\frac{\dot{f}(t)}{2H}+g(t)\right)\, , \label{eq:Ca11} 
\end{align} 
with the functions $f(t)$ and $g(t)$ playing the roles of the entropic terms. For the so-called $\Lambda$(t) models, assuming $f(t) = g(t)$, so that we have
\begin{equation}\label{f(t)}
f(t) = \frac{8 \pi G}{3} \rho_{e}.
\end{equation}
The continuity equations will read as
\begin{align}
\sum_i \dot{\rho}_i + 3 H \sum \rho_i (1+w_i) = -\dot{\rho}_e, \label{eq:continuity_Lambda_MLn_1}
\end{align} 
where barotropic equation of state, $p_i = w_ic^2\rho_i$ for matter $(w_m=0)$ and radiation $(w_r=1/3)$ have been used. From Eq. (\ref{rho_n}), one can write
\begin{equation}
\dot{\rho}_{e} = C_e\,H \dot{H}, \label{rhondot}
\end{equation}
with $C_e= 2\gamma/(4 \pi G)$. From Eqs.~(\ref{eq:Fa11}) and (\ref{eq:Ca11}) one can calculate 
\begin{equation}\label{hdot}
\dot{H}  = -4\pi G \left(\rho_m + \frac{4}{3} \rho_r\right)
\end{equation}
and using it in Eq. (\ref{rhondot}), one gets
\begin{align}\label{eq:continuity_Lambda_MLn_1_entropy}
\dot{\rho}_{e}=
-A_{m} H\rho_m - A_{r} H \rho_r,
\end{align}
where $A_{m} = 4\pi G\, C_e$ and $A_{r} = 16\pi G/3\,C_e$. Using Eq. (\ref{eq:continuity_Lambda_MLn_1_entropy}) one can further separate Eq.~(\ref{eq:continuity_Lambda_MLn_1}) for matter and radiation,
\begin{align}
&a \rho'_i + 3 \rho_i (1+w_{eff,i}) = 0 \,, \label{eq:continuity_Lambda_MLn_mat_rad} \\
&w_{eff,i} = w_i - \frac{A_{i}}{3}\, . \label{eq:weff_1}
\end{align}

By substituting equation \eqref{f(t)} into equation \eqref{eq:Fa11}, and then using equation \eqref{eq:continuity_Lambda_MLn_1_entropy}, we obtain
\begin{align}
H(a,\gamma)&= H_0\!\bigg[ 
      \Omega_{m,0}\,a^{-3\alpha}
    + \Omega_{r,0}\,a^{-4\alpha}
    + \Omega_{e,0} \notag \\
& -\frac{2\gamma\,\Omega_{m,0}}{-3\alpha}\bigl(a^{-3\alpha}-1\bigr)
  -\frac{8\gamma\,\Omega_{r,0}}{-12\alpha}\bigl(a^{-4\alpha}-1\bigr)
\bigg]^{1/2},                                         \label{eq:Hofagam}
\end{align}
where $\alpha=\left(1-\frac{2\gamma}{3}\right)$, $H_0 = 100\,h$ km/s/Mpc, $\Omega_{m,0}$ and $\Omega_{r,0}$ are the present-day matter and radiation density parameters, respectively, and $\Omega_{\mathrm{e},0} = 1 - \Omega_{m,0} - \Omega_{r,0}$.
\begin{figure}[h!t]
    \centering
    \includegraphics[width=1\linewidth]{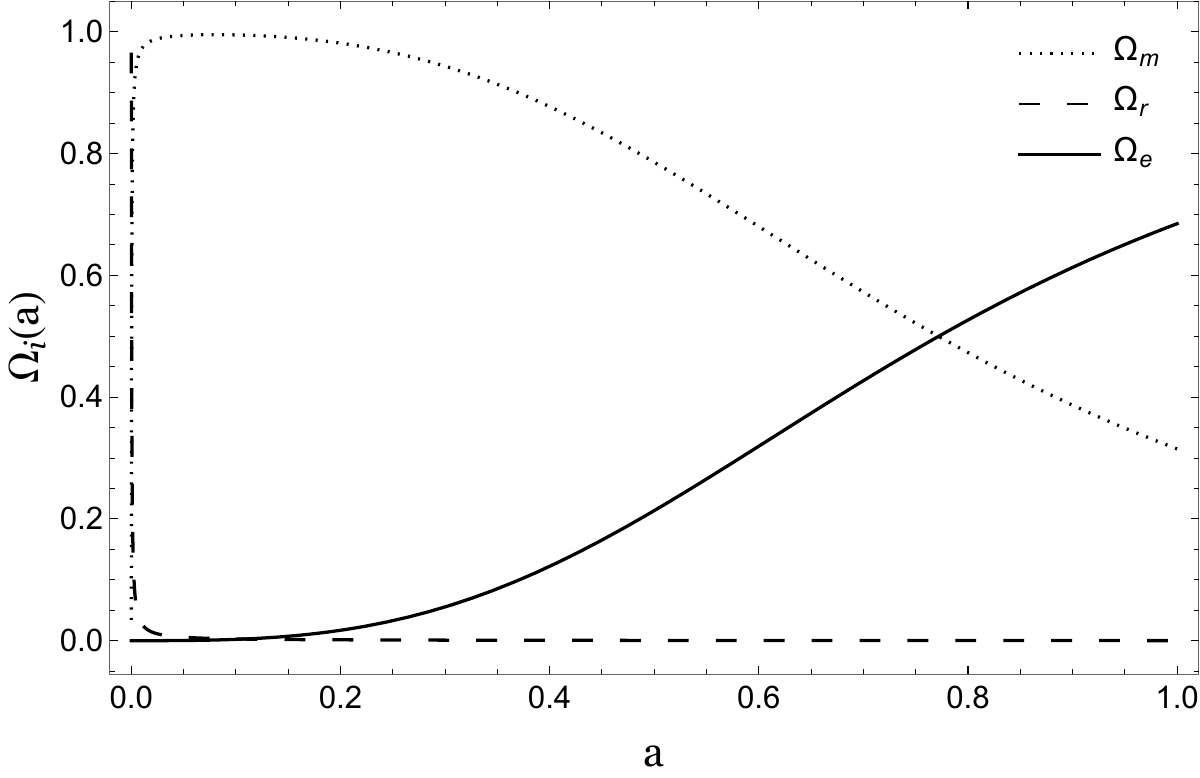}
    \caption{Evolution of the normalized density parameters $\Omega_{\mathrm{m}}(a)$, $\Omega_{\mathrm{r}}(a)$ and $\Omega_{\mathrm{e}}(a)$ as functions of $a$. 
    }
    \label{fig:omegas}
\end{figure}
In Fig.~\ref{fig:omegas} the evolution of the normalized energy densities $\Omega_m(a)$, $\Omega_r(a)$, $\Omega_e(a)$ are shown as a function of the scale factor (for reference value $\gamma=10^{-5}$ chosen as already reproducing $\Lambda$CDM behaviour, see below). 

\begin{figure}[h!t]
    \centering
    \includegraphics[width=1\linewidth]{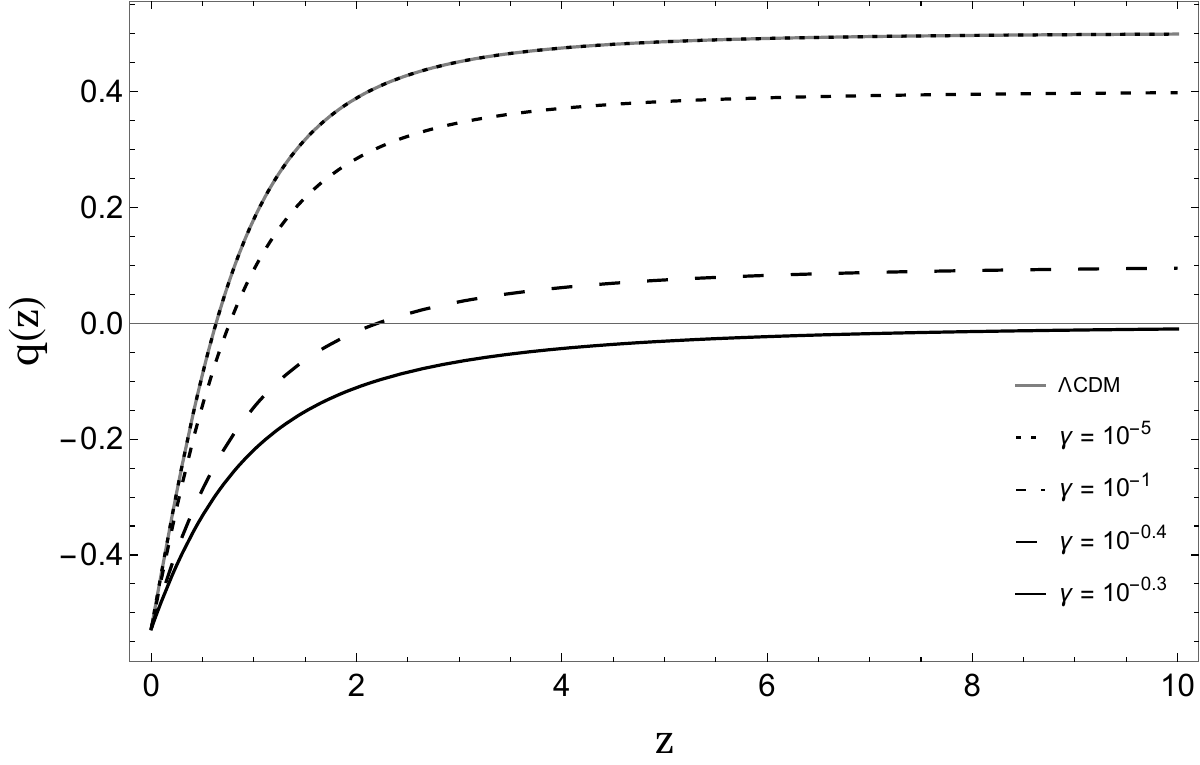}
    \caption{Comparison of the deceleration parameter \( q(z) \) for \(\Lambda\)CDM and MHEC with different values of \( \gamma \).}
    \label{fig:qz}
\end{figure}
Using best-fit value of $h =0.68$ and $\gamma=10^{-5}$ we plot the deceleration parameter $q(z)$ - Fig. \ref{fig:qz}. The trajectory traces the $\Lambda$CDM history, with  the deceleration–to–acceleration transition at $z_{tr}\simeq 0.63$. Background solutions with larger $\gamma$ makes entropic term dominate ealier. Increasing $\gamma$ shifts the transition to larger redshift.

\begin{figure}[h!t]
    \centering
    \includegraphics[width=1\linewidth]{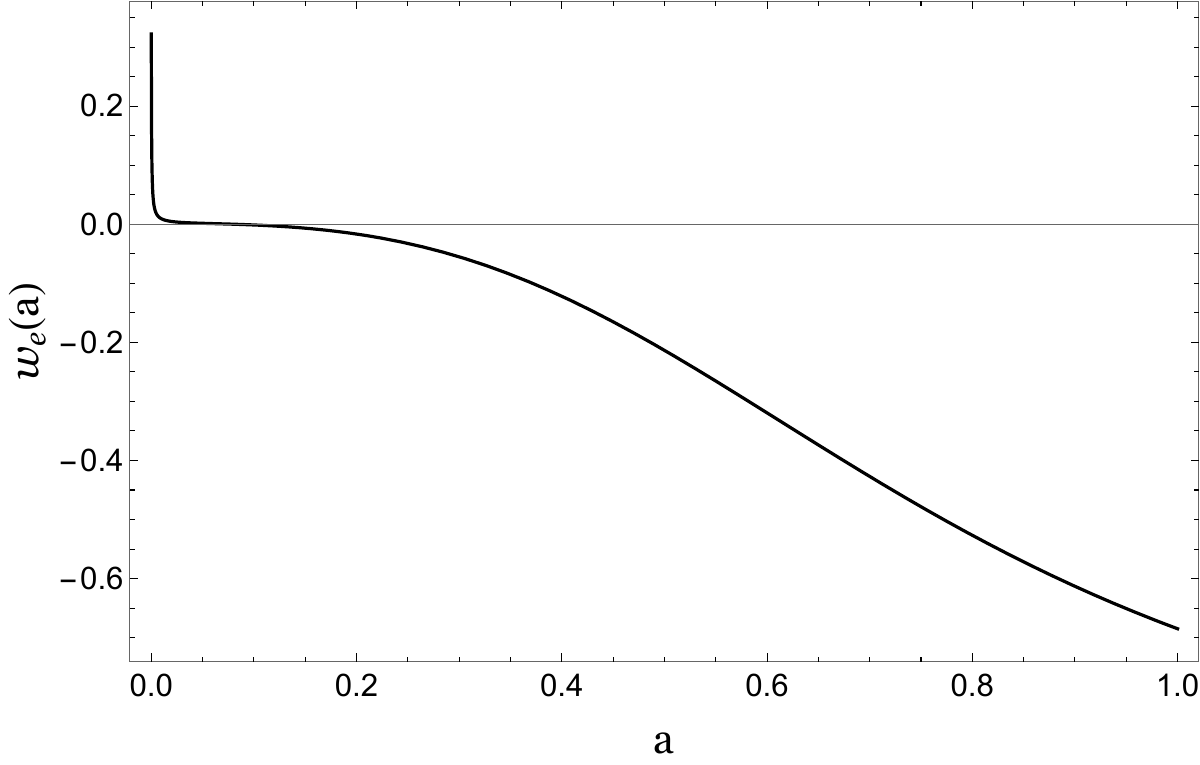}
    \caption{Evolution of the equation of state \( w_{\mathrm{e}}(a) \) as a function of the scale factor \( a \).}
    \label{fig:we}
\end{figure}

Figure~\ref{fig:we} illustrates the scale–factor evolution of the effective entropic equation-of-state parameter \(w_{e}(a)\).  
At early times (\(a\!\lesssim\!0.2\)) the entropic component acts as radiation, then as a dust \(w_{e}\simeq 0\), as the Universe enters the matter era \(w_{e}\) steadily declines, crosses the “quintessence divide’’ \(w=-1/3\) around \(a\simeq0.55\,(z\simeq0.8)\), and reaches less than \(w_{e}\approx-0.6\) today. 
This smooth, monotonic transition—from a mildly positive value to a negative, acceleration-driving regime—mirrors the behaviour of the deceleration parameter in Fig.~\ref{fig:qz}  and encapsulates how the same entropic term can act like an extra radiation component in the early Universe while fuelling late-time cosmic acceleration.

\begin{figure}[h!t]
    \centering
    \includegraphics[width=1\linewidth]{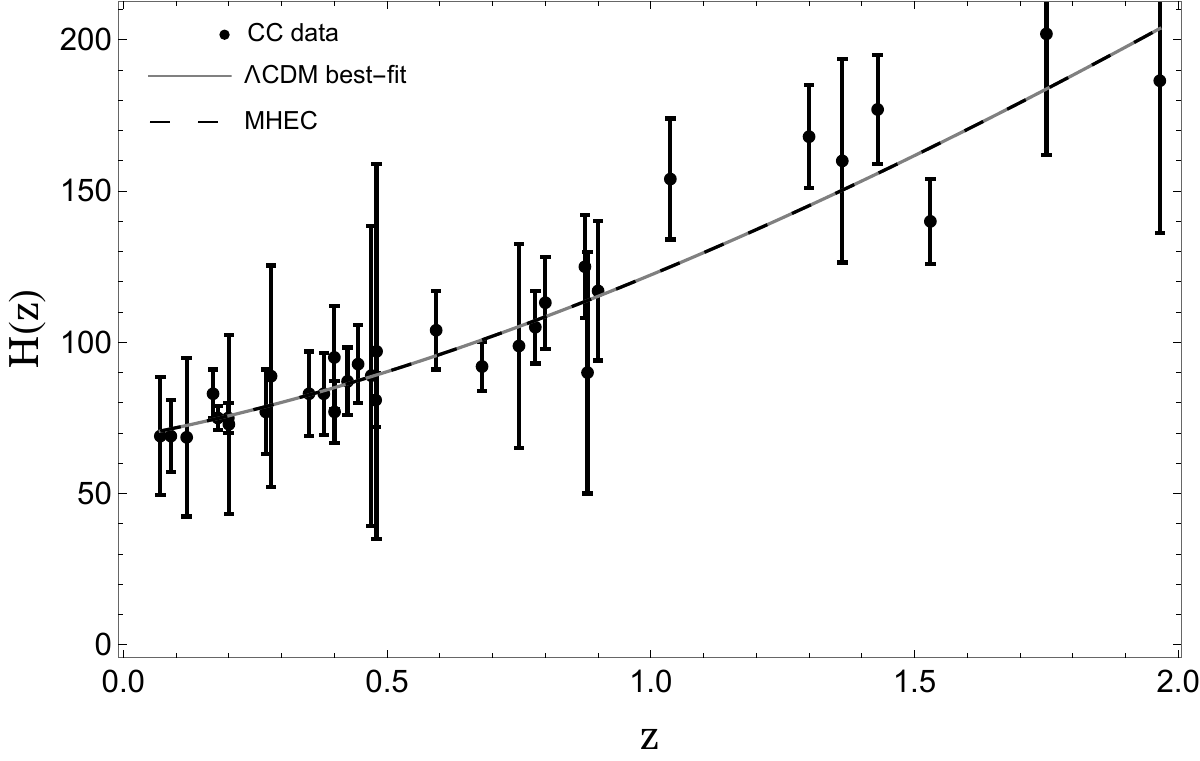}
    \caption{Hubble parameter data from cosmic chronometers (CC) with best-fit theoretical predictions for $\Lambda$CDM and MHEC with $\gamma=10^{-5}$. The \textbf{black} points with error bars represent the CC data. The solid \textbf{gray} curve is the best-fit $H(z)$ evolution of the standard $\Lambda$CDM model; the dashed \textbf{black} curve represents MHEC.}
    \label{fig:hz}
\end{figure}

Having thus established that the entropic component evolves smoothly from radiation-like to acceleration-driving behaviour, we now confront the same background solution with direct measurements of the expansion rate. Figure \ref{fig:hz} shows the resulting comparison with the cosmic-chronometer $H(z)$ data \cite{Zhang:2012mp,Simon:2004tf,Moresco:2012jh,BOSS:2016wmc,Moresco:2016mzx,Ratsimbazafy:2017vga,Stern:2009ep,Jiao:2022aep,Moresco:2015cya}, on which we base the statistical fit described below.


We performed a detailed statistical fit to observational data for the Hubble parameter $H(z)$ as measured by cosmic chronometers (CC). We fixed $\Omega_{m,0}=0.315$ and allowed two parameters $h$ and $\gamma$ to vary freely within physical priors: $0.5 \leq h \leq 0.9$, and $0\le\gamma \leq 10^{-2}$.

For the statistical comparison between theory and data, we computed the chi-squared statistic as
$$
\chi^2(h, \gamma) = \Delta \mathbf{H}^T \, \mathbf{C}^{-1} \, \Delta \mathbf{H}
$$
where \( \mathbf{C}^{-1} \) is the full inverse covariance matrix provided for the CC dataset.

The minimization of the chi-squared was carried out using the  global optimization method. The best-fit parameter values were identified at the minimum chi-squared.
The resulting best fit values are found to be $h = 0.68 \pm 0.02$, and the $\gamma$ has only an upper bound $\gamma<0.02$ at (95\% CL for fixed best fit value of $h$). CC data leave $\gamma$ unconstrained and tighter limits must come from growth or high-$z$ probes. Values near
$\gamma\sim10^{-5}$ reproduce the $\Lambda$CDM background almost exactly -- see Fig.~\ref{fig:hz}, and
even $\gamma\sim10^{-2}$ shifts the $H(z)$ curve by less than the current CC
uncertainties. These background results confirm that the entropic scenario -- see also Ref.\cite{Gohar2024} -- matches all distance-based probes, but the decisive test is whether it also reproduces the observed growth of cosmic structures—a question we tackle quantitatively in Sec.~\ref{sec:discussion}.
\section{Perturbations}\label{sec:perturbations}
\subsection{General set–up}\label{subsec:setup}
We work on the spatially–flat FLRW background
\begin{equation}
ds^{2} = -dt^{2} + a^{2}(t)\,\delta_{ij}\,dx^{i}dx^{j},
\end{equation}
filled with pressureless matter ($w_{m}=0$), radiation ($w_{r}=1/3$) and a
homogeneous dark-energy component that does \emph{not} fluctuate in the cases considered below.  
Cosmic time derivatives are denoted by overdots and comoving derivatives by a vertical bar,
$\dot{f}\equiv\partial f/\partial t$, $f_{|i}\equiv\partial f/\partial x^{i}$.

Throughout we follow the Newtonian (sub-Hubble) treatment developed in
Refs.\,\cite{Lima1997,Jesus2011,Reis2003,Nayeri1998,Koivisto2011,Komatsu2014} and used extensively in 
interacting and entropic cosmologies
\cite{,Komatsu2014,Komatsu2014a,Basilakos2014,SolaPeracaula2018}.
Proper coordinates $\vec r=a\vec x$ are related to the comoving ones $\vec x$ in the usual way,
and we employ the gauge where the background 4-velocity is
$u^\mu_{\text{bg}}=(1,\vec 0)$.

\subsection{Governing equations}
For each fluid component $i\in\{m,r\}$ the Euler, Poisson and Continuity equations in proper coordinates are
\begin{align}
\frac{\partial\vec u_i}{\partial t}+(\vec u_i\!\cdot\!\nabla_r)\vec u_i
      &= -\nabla_r\Phi_i-\frac{\nabla_r P_i}{\rho_i+P_i},\label{eq:euler_proper}\\
\nabla_r^2\Phi_i&=4\pi G\sum_j (\rho_j+3P_j),\label{eq:poisson_proper}\\
\frac{\partial\rho_i}{\partial t}+\nabla_r\!\cdot\!(\rho_i\vec u_i)
      +P_i\,\nabla_r\!\cdot\!\vec u_i &=2\gamma\,H\,(1+w_i)\rho_i.
      \label{eq:cont_proper}
\end{align}

A first-order splitting into background and perturbations,
\begin{align}
\rho_i       &=\bar\rho_i(t)\bigl[1+\delta_i(\vec r,t)\bigr],\qquad
P_i=\bar P_i+\delta P_i,\\
\vec u_i     &=\vec 0+\vec v_i,\qquad
\Phi_i       =\bar\Phi_i+\varphi_i,
\end{align}
together with the transformation rules
$\nabla_r=a^{-1}\nabla_x$ and
$\partial_t|_{\vec x}=\partial_t|_{\vec r}+(\dot a/a)(\vec x\!\cdot\!\nabla_x)$
yields, after linearisation,
\begin{align}
\dot{\vec v}_i+\frac{\dot a}{a}\vec v_i
  &= -\frac{1}{a}\nabla_x\varphi_i
     -\frac{1}{a}\frac{\nabla_x\delta P_i}{\bar\rho_i+\bar P_i},\label{eq:euler_com}\\
\nabla_x^2\varphi_i
  &= 4\pi Ga^{2}\sum_j\bigl(\bar\rho_j+3\bar P_j\bigr)\delta_j.\label{eq:poisson_com}
\end{align}
With the definitions
$c_{\mathrm{eff}}^{2}\equiv\delta P_i/\delta\rho_i$ and $w_i\equiv\bar P_i/\bar\rho_i$
\eqref{eq:euler_com} becomes
\begin{equation}\label{eq:euler_final}
\dot{\vec v}_i+\frac{\dot a}{a}\vec v_i
+\frac{1}{a}\nabla_x\varphi_i
+\frac{c_{\mathrm{eff}}^{2}}{a(1+w_i)}\nabla_x\delta_i
=0.
\end{equation}
Similarly, Eq.,\eqref{eq:poisson_com} rewrites as
\begin{equation}\label{eq:poisson_final}
\nabla_x^2\varphi_i = 4\pi G a^{2} \sum_j \bigl(1+3c_{\mathrm{eff},j}^{2}\bigr)\bar\rho_j\delta_j.
\end{equation}
Linearising the continuity equation \eqref{eq:cont_proper} gives  
\begin{equation}\label{eq:cont_lin}
\dot{\delta}_i + \frac{1+w_i}{a}\,\nabla_x\!\cdot\!\vec v_i
      = 2\gamma H(1+w_i)\,\delta_i
        - \frac{\widehat Q_i}{\bar\rho_i}\;,
\end{equation}
where $\widehat Q_i\equiv Q_i-\bar Q_i =
      2\gamma H(1+w_i)\bar\rho_i\,\delta_i$ is the first–order part of the
interaction source.  Two choices are common in the literature:
(i)~\emph{perturbed interaction kept},
$\widehat Q_i = 2\gamma H(1+w_i)\bar\rho_i\,\delta_i$,
which cancels the right–hand side of \eqref{eq:cont_lin};
(ii)~\emph{perturbed interaction neglected},
$\widehat Q_i = 0$, which leaves the drag term
$2\gamma H(1+w_i)\delta_i$ explicit.
These alternatives generate the
“Case A” and “Case B” systems written below.
\subsection{Evolution of the density contrasts}
\begin{enumerate}[
                    style=nextline,
                    label=\textnormal{Case \Alph*:},
                    font=\normalfont,
                    leftmargin=0.pt,      
                    itemindent=\parindent,
                    align=left]       
\item interaction \emph{perturbations included.}\\
Taking the divergence of \eqref{eq:euler_final}, substituting the resulting expression along with \eqref{eq:poisson_final} into the derivative of perturbed continuity equation $\dot{\delta}_i+\frac{1+w_i}{a}\nabla_x\!\cdot\!\vec v_i=0$, and applying a Fourier transform using the convention $\nabla_x^{2} \delta_i = -k^{2} \delta_i$, we arrive at 
\begin{equation}\label{eq:master}
\ddot{\delta}_i + 2H\dot{\delta}_i
      - 4\pi G\,(1+w_i)\!\!\sum_j (1+3w_j)\,\bar\rho_j\,\delta_j
      + \frac{c_{\mathrm{eff}}^{2}k^{2}}{a^{2}}\,\delta_i = 0 .
\end{equation}
With $c_{\mathrm{eff}}^{2}=w_i$ (adiabatic perturbations) and setting $\delta\rho_{\text{e}}=0$ we get the coupled system

\begin{align}
\ddot\delta_m+2H\dot\delta_m
   &=4\pi G\bigl(\bar\rho_m\delta_m+2\bar\rho_r\delta_r\bigr),\\
\ddot\delta_r+2H\dot\delta_r
   &=\frac{16}{3}\pi G\bigl(\bar\rho_m\delta_m+2\bar\rho_r\delta_r\bigr)
     +\frac{k^{2}}{3a^{2}}\delta_r.
\end{align}

\item interaction \emph{perturbations neglected.}\\
If the interaction source terms couple only to the homogeneous densities, taking the divergence of \eqref{eq:euler_final}, substituting the resulting expression along with \eqref{eq:poisson_final} into the derivative of perturbed continuity equation  
$\dot{\delta}_i + \frac{1+w_i}{a}\,\nabla_x\!\cdot\!\vec v_i
      = 2\gamma H(1+w_i)\,\delta_i$, and applying a Fourier transform using the convention $\nabla_x^{2} \delta_i = -k^{2} \delta_i$, we arive at
\begin{align}
&\ddot{\delta}_i + 2H\dot{\delta}_i
      - 4\pi G\,(1+w_i)\!\!\sum_j (1+3w_j)\,\bar\rho_j\,\delta_j
      + \frac{c_{\mathrm{eff}}^{2}k^{2}}{a^{2}}\,\delta_i\notag\\ &+4\gamma H^2\delta_i(w_i+1)+\frac{\partial}{\partial t}\left[2\gamma H\delta_i(w_i+1)\right]= 0 .
\end{align}

With $c_{\mathrm{eff}}^{2}=w_i$ (adiabatic perturbations) and $\delta\rho_{\text{e}}=0$ we get the coupled system
\begin{align}
\ddot\delta_m+2H(1+\gamma)\dot\delta_m
   &+4\gamma H^{2}\delta_m+2\gamma(\dot H)\delta_m\nonumber\\
   &=4\pi G\bigl(\bar\rho_m\delta_m+2\bar\rho_r\delta_r\bigr),\\
\ddot\delta_r+2H(1+\tfrac{4}{3}\gamma)\dot\delta_r
   &+\tfrac{16}{3}\gamma H^{2}\delta_r
     +\tfrac{8}{3}\gamma\dot H\,\delta_r
     +\frac{k^{2}}{3a^{2}}\delta_r\nonumber\\
   &=\frac{16}{3}\pi G\bigl(\bar\rho_m\delta_m+2\bar\rho_r\delta_r\bigr).
\end{align}

These equations form the basis for the numerical analysis in Refs.\,\cite{GomezValent2017,Komatsu2014} and can be specialised to pressureless matter alone ($w_m=0$) for late-time calculations of $f\sigma_{8}$ or the matter power spectrum.
\end{enumerate}
\subsection{Perturbing or not perturbing the interaction source}
\label{subsec:deltaQ_choice}

The linear system derived above splits into two distinct regimes depending on whether the interaction four-vector $Q_\mu$ is perturbed together with the densities:

\medskip
\noindent
\textit{Case A — perturbations included.}
Here one keeps $\widehat Q_i = Q_i\,\delta_i$ in the perturbed continuity equation.  The $\gamma$-dependent contribution that appears on the right-hand side of the background equation is then exactly multiplied by the same density contrast and cancels against the explicit drag term.  As a result,
the coupled matter–radiation system written immediately above for \emph{Case A} contains no trace of $\gamma$ beyond its effect on the background functions $H(a)$ and $\bar\rho_i(a)$.  Linear growth therefore follows the standard $\Lambda$CDM form, modified only indirectly through the altered expansion history.

\medskip
\noindent
\textit{Case B — perturbations neglected.}
If, instead, one sets $\widehat Q_i = 0$, the drag produced by the background
exchange survives and appears explicitly in the perturbation equations
shown for \emph{Case B}.  
Two new pieces are generated: an extra friction term, proportional to $2\gamma H(1+w_i)\dot\delta_i$, and an effective mass shift, $\left[4\gamma H^{2}+2\gamma\dot H\right](1+w_i)
\delta_i$. Both contribute at $\mathcal{O}(\gamma)$ and introduce a scale–independent tilt in the growth index that is absent in Case A.

\medskip
\noindent
\textit{Choice adopted in this work.}
Because local energy–momentum conservation requires a consistent treatment of $Q_\mu$ at the perturbative level, all numerical results that follow are obtained with the \emph{Case A} equations.  Readers interested in the background–only prescription can reproduce it by replacing the growth
system with the \emph{Case B} set already given above, but should bear in mind that this approximates—and generally amplifies—the impact of~$\gamma$ on structure formation.
\subsection{Remarks}\label{sec:remarks}
\begin{itemize}
\item The $\,H$– and $\,\dot H$–terms induced by entropic or running-vacuum scenarios enter only through the background $H(t)$ and its derivative; at \emph{linear} order they never introduce explicit scale dependence beyond the usual Jean’s term $\propto k^{2}$ \cite{Koivisto2011,Basilakos2014,GomezValent2017}.  
\item Including or neglecting the perturbations in the interaction four-vector $Q_\mu$ leads to qualitatively different damping terms $\propto\gamma$ in the matter equation, cf. Case A vs.\ Case B. This is the main driver of the growth-rate split originates from the extra damping term that appears only when the interaction perturbation is neglected, as already emphasized in Refs.,\cite{Koivisto2011,Jesus2011,Basilakos2014,GomezValent2017,SolaPeracaula2018}. 
\end{itemize}
\section{Discussion}\label{sec:discussion}
Having verified in Sec.\ref{subsec:bg-plots} that the entropic model reproduces background distance indicators, we now turn to its linear-growth sector and spell out the quantities entering our $f\sigma_8$ analysis.

The comoving angular diameter distance is
\begin{equation}
D_A(z) = \frac{1}{1+z} \int_0^z \frac{c}{H(z')} \, dz',
\label{eq:DA}
\end{equation}
where $c$ is the speed of light.

The growth factor $D(a)$ is given by
\begin{equation}
D(a) = \frac{\delta_M(a)}{\delta_M(a=1)}.
\end{equation}
The linear growth rate is
\begin{equation}
f(a) = \frac{d\ln D}{d\ln a}
\end{equation}
and 
$$\sigma_{8}(a)=\sigma_{80} D(a)
$$
\begin{equation}
f\sigma_8(a) = a \sigma_{80} D'(a).
\end{equation}

To compare the theoretical prediction to each $f\sigma_8(a)$ measurement, we rescale using the fiducial cosmology. For each data point,
\begin{equation}
f\sigma_8^{\mathrm{mr}}(z) = f\sigma_8^{\mathrm{model}}(z) \frac{H_{\mathrm{fid}}(z)\, D_{A,\,\mathrm{fid}}(z)}{H(z)_{model}\, D_A(z)_{model}},
\end{equation}
where $f\sigma_8^{\mathrm{mr}}$ stands for the model rescaled. 
Chi-squared is defined as
\begin{equation}
\chi^2(\gamma, \sigma_{80}) = \sum_{i}
\left[ \frac{
f\sigma_8^{\mathrm{mr}}(z_i) - f\sigma_8^{\mathrm{obs}}(z_i)
}{\sigma_i} \right]^2
\end{equation}
where $f\sigma_8^{\mathrm{obs}}(z_i)$ and $\sigma_i$ are the observed value and uncertainty at each redshift $z_i$ - the data are taken from \cite{Song2009,Achitouv2017,Okumura2016,Blake2013,Marin2015,Bhattacharyya2021,Mattia2020,Chapman2022,Howlett2015,Hawken2017,Mohammad2018,Jullo2019}.

We fixed the matter density to $\Omega_{m,0}=0.315$ and adopted 
$h=0.68$ from the $H(z)$ background fit, while allowing the entropic 
coupling and the present-day normalisation of fluctuations to vary over 
the physical ranges
$0\le\gamma\le10^{-2}$ and $0.5\le\sigma_{8,0}\le1$.  
The best-fit values of $(\gamma,\sigma_{8,0})$ were obtained by 
minimising the $\chi^{2}$ function.

\begin{figure}[h!t]
    \centering
    \includegraphics[width=1\linewidth]{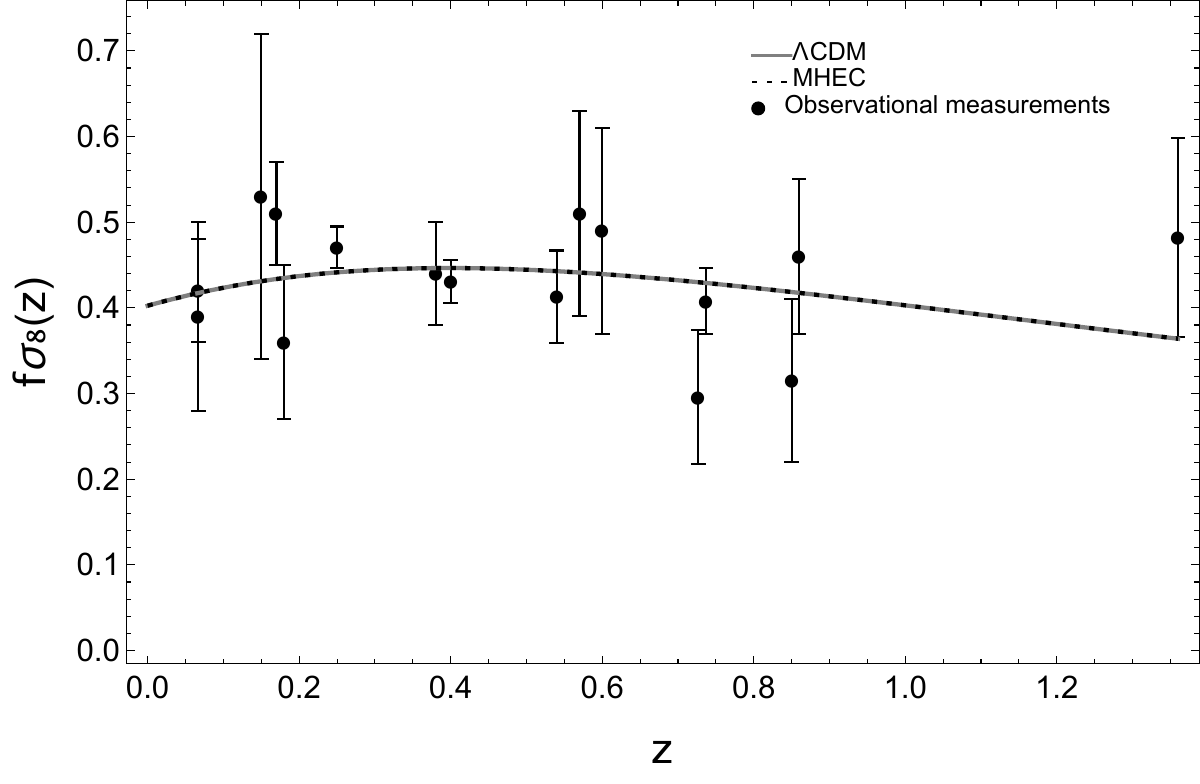}
    \caption{$f\sigma_8(z)$ data compared with $\Lambda$CDM and MHEC -- $\gamma=10^{-5}.$}
    \label{fig:fs8}
\end{figure}

Fig.~\ref{fig:fs8} displays the growth measurements alongside two theoretical curves: the reference \(\Lambda\)CDM prediction (\(\gamma=0\)) and an entropic model with \(\gamma=1\times10^{-5}\).  The global minimum of the $\chi^{2}$ surface actually sits at $\gamma\simeq0$, $\sigma_{8,0}=0.76$, i.e.\ exactly the $\Lambda$CDM limit; however, a value as small as $\gamma=10^{-5}$ is already indistinguishable from that optimum within the current error bars, so we plot it as a representative entropic curve. The two lines overlap to within the plotting resolution, emphasising that used $f\sigma_{8}$ data cannot distinguish the entropic scenario from standard $\Lambda$CDM for \(\gamma\lesssim10^{-5}\).

\begin{figure}[h!]
    \centering
 \includegraphics[width=1\linewidth]{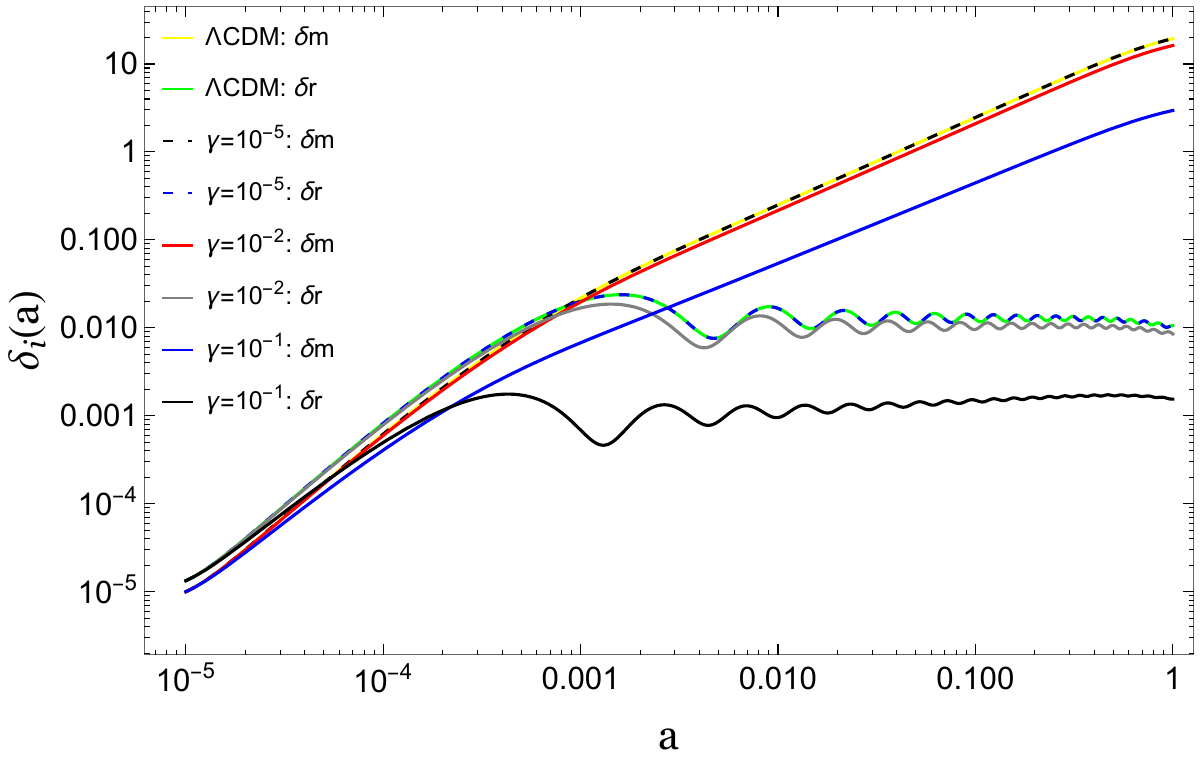}
    \caption{The evolution of $\delta_m$ and $\delta_r$ density contrasts as functions of the scale factor~$a$, for $\Lambda$CDM and MHEC.}
    \label{fig:deltaCA}
\end{figure}

\paragraph{Perturbation behaviour.}
Fig.~\ref{fig:deltaCA} (Case--A) confirms the expectation drawn from the background analysis: when the interaction term is \emph{perturbed} consistently, the entropic coupling feeds into structure formation only through the modified expansion rate.  As a result the growth histories for
$\gamma=10^{-5}$ and even $\gamma=10^{-2}$ lie on top of the $\Lambda$CDM curves down to the line thickness; a visible departure appears for the extreme choice $\gamma=10^{-1}$.

\begin{figure}[h!]
    \centering
 \includegraphics[width=1\linewidth]{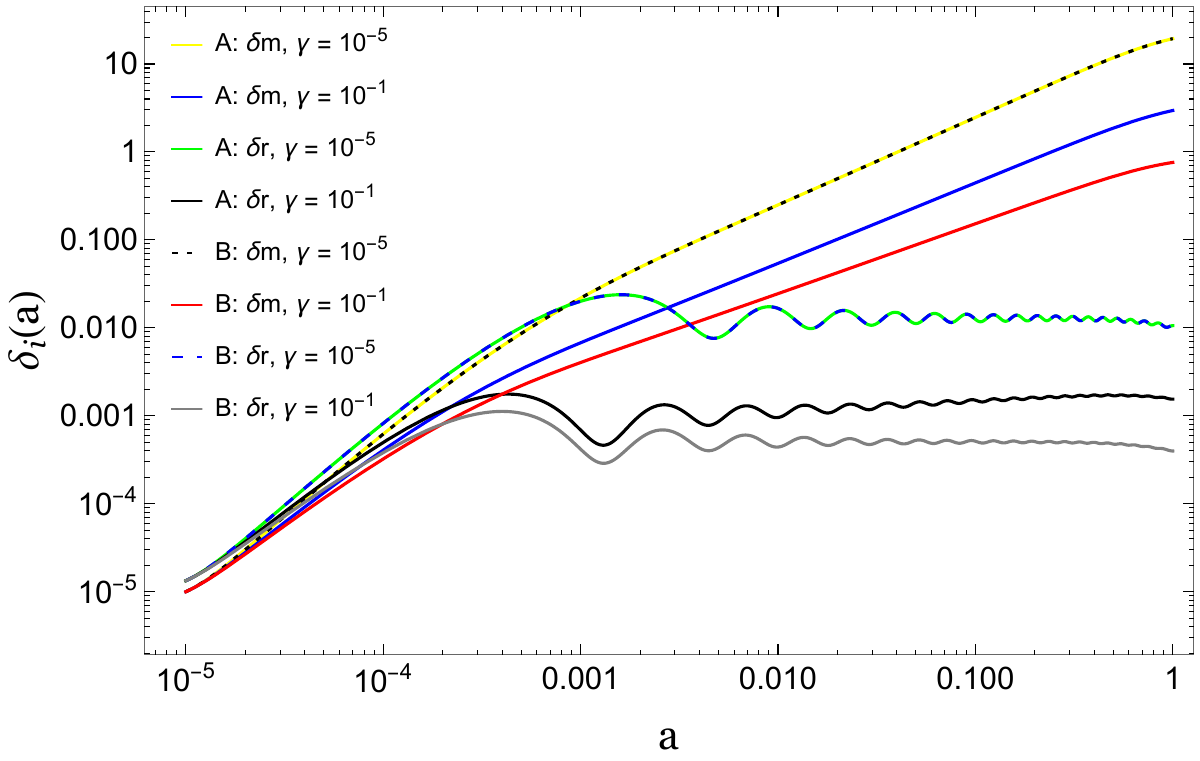}
    \caption{The evolution $\delta_m$ and $\delta_r$ as functions of the scale factor~$a$, for cases A and B.}
   \label{fig:deltaCAB}
\end{figure}

\begin{figure}[h!t]
    \centering
    \includegraphics[width=1\linewidth]{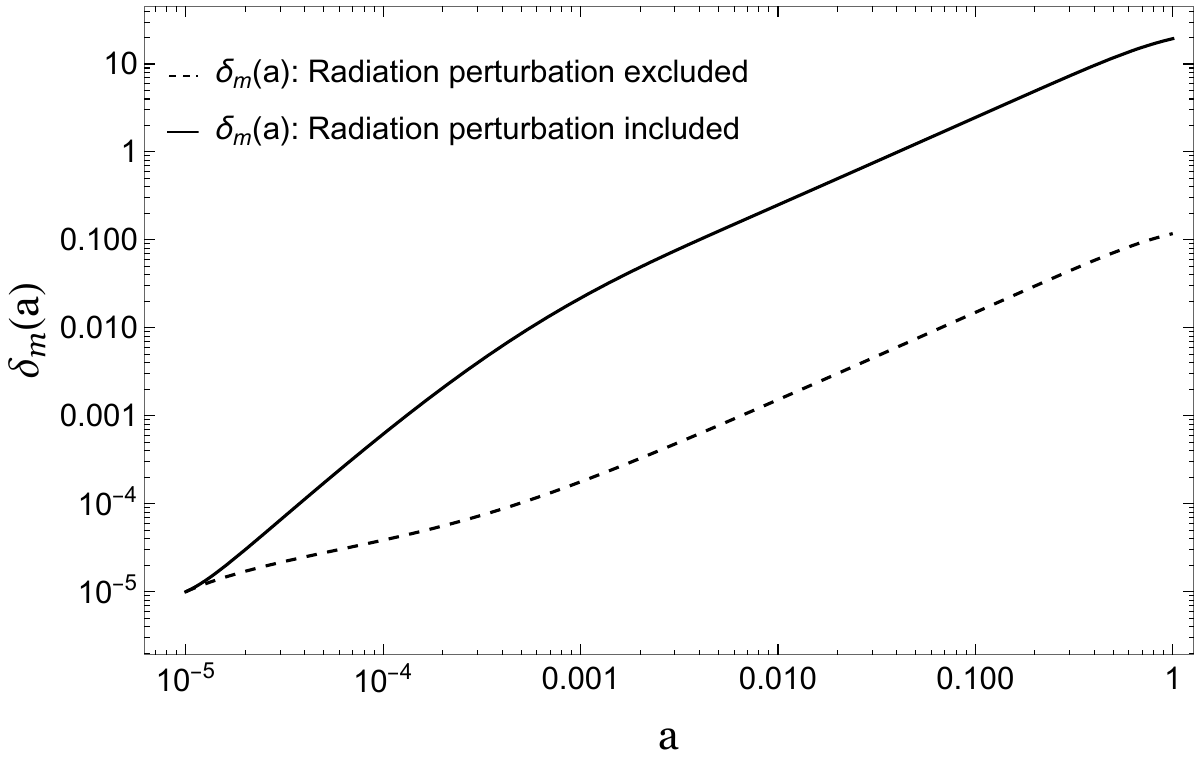}
    \caption{Effect of radiation perturbations on matter growth in Case~A.}
    \label{fig:deltaReffect}
\end{figure}

\paragraph{Importance of radiation perturbations.}
Before contrasting Case~A with Case~B it is useful to isolate the role of the radiation mode itself.  
Fig.~\ref{fig:deltaReffect} shows two Case--A integrations that start from the same initial conditions
the solid curve evolves the full coupled system \(\{\delta_{m},\delta_{r}\}\),  
while the dashed curve repeats the calculation after setting
\(\delta_{r}=0\). The feedback from \( \delta_r \) boosts the growth of \( \delta_m \) by over 100 times the value obtained by setting \( \delta_r = 0 \). It demonstrates that a
\emph{consistent} treatment of radiation is mandatory when one wishes to
propagate the system from high redshift or to interface with CMB.

Fig.~\ref{fig:deltaCAB} compares this “conservative’’ treatment with Case~B, in which the perturbation in the interaction four--vector is \emph{switched off}.  The additional friction and effective--mass terms introduced in Eqs.\,\eqref{eq:continuity_Lambda_MLn_1}–\eqref{eq:weff_1} now act directly on the density modes and suppress their growth. For \( \gamma = 10^{-1} \), the matter amplitude \( \delta_m \) in Case B is suppressed by a factor of approximately $\sim 4$ relative to Case A at \( a = 1 \). 

\medskip
\noindent
\textit{Implications.}  Taken together, Figs.~\ref{fig:fs8}, \ref{fig:deltaCA} and \ref{fig:deltaCAB} demonstrate that the thermodynamically consistent, fully perturbed version of the MHEC model (Case A) can be fitted to present growth data as successfully as $\Lambda$CDM, whereas ad-hoc choices that neglect the interaction perturbation run into observational conflict already for smaller values of $\gamma$. 
This robustness against structure–formation tests, achieved without introducing additional free parameters, underscores the viability of the entropic scenario and motivates a full MCMC analysis, a task now in progress. 
\section{Conclusions}\label{sec:conclusions}
The generalized mass–to–horizon–relation entropic dark–energy model presented in this work satisfies the two most delicate requirements that have historically challenged entropic–force cosmologies, namely internal thermodynamic consistency and compatibility with the observed growth of cosmic structures.  Our discussion has dealt exclusively with the linear mass–to–horizon relation of Eq.~(\ref{mhr}); together with the associated entropy of Eq.~(\ref{genentropy}) and the standard Hawking temperature of \(T_{\mathrm H}= \hbar c /(2\pi k_{\mathrm B}L)\), this choice corresponds to the case \(n=1\).

Because the pair \((M(L),S(L))\) obeys exactly the first–law identity \(dE=c^{2}\,dM=T_{\mathrm H}\,dS_{n}\), the Clausius relation is preserved.  Since the entropy remains proportional to the Bekenstein–Hawking value, all Legendre relations—including Euler homogeneity and the Gibbs–Duhem equation—are intact, providing a closed thermodynamic description on the horizon.  The background dynamics can be written in the $\Lambda$CDM form \(H^{2}=H_{0}^{2}\,[\Omega_{m0}(1+z)^{3}+\Omega_{e}]\), where \(\Omega_{e}\) represents the constant part of the entropic density; a combined fit to SNIa, BAO, CMB distance priors and cosmic–chronometer data gives \(\Omega_{e}=0.684\pm0.012\) with a goodness of fit indistinguishable from the concordance model \cite{Gohar2024}.

At the perturbative level the effective gravitational coupling differs from Newton’s constant. Previous entropic models failed either because they relied on non-extensive entropies incompatible with the Hawking temperature or because they lacked an additive vacuum term and consequently suppressed structure growth by more than forty per cent \cite{Basilakos2014,Gohar2024}.  The present \(n=1\) framework avoids both pitfalls: the entropy is extensive and proportional to \(S_{\mathrm{BH}}\), ensuring thermodynamic closure, while the constant \(\Omega_{e}\) component restores the correct deceleration–to–acceleration transition and leaves the canonical growth history virtually unchanged.

We therefore conclude that the generalized mass-to-horizon relation entropic cosmology provides a self-consistent thermodynamic scheme that is fully compatible with current background and growth observations, refuting earlier claims that entropic–force cosmologies cannot reproduce the observed structure-formation rate.

Having shown that the MHEC model matches the current $f\sigma_8$ measurements as well as $\Lambda$CDM the next step is a dedicated, full-MCMC analysis of the growth data -- allowing the entropic parameters to vary -- a study we are now undertaking.
\bibliographystyle{apsrev4-2}
\bibliography{ali_perts}

\begin{thebibliography}{57}%
\makeatletter
\providecommand \@ifxundefined [1]{%
 \@ifx{#1\undefined}
}%
\providecommand \@ifnum [1]{%
 \ifnum #1\expandafter \@firstoftwo
 \else \expandafter \@secondoftwo
 \fi
}%
\providecommand \@ifx [1]{%
 \ifx #1\expandafter \@firstoftwo
 \else \expandafter \@secondoftwo
 \fi
}%
\providecommand \natexlab [1]{#1}%
\providecommand \enquote  [1]{``#1''}%
\providecommand \bibnamefont  [1]{#1}%
\providecommand \bibfnamefont [1]{#1}%
\providecommand \citenamefont [1]{#1}%
\providecommand \href@noop [0]{\@secondoftwo}%
\providecommand \href [0]{\begingroup \@sanitize@url \@href}%
\providecommand \@href[1]{\@@startlink{#1}\@@href}%
\providecommand \@@href[1]{\endgroup#1\@@endlink}%
\providecommand \@sanitize@url [0]{\catcode `\\12\catcode `\$12\catcode `\&12\catcode `\#12\catcode `\^12\catcode `\_12\catcode `\%12\relax}%
\providecommand \@@startlink[1]{}%
\providecommand \@@endlink[0]{}%
\providecommand \url  [0]{\begingroup\@sanitize@url \@url }%
\providecommand \@url [1]{\endgroup\@href {#1}{\urlprefix }}%
\providecommand \urlprefix  [0]{URL }%
\providecommand \Eprint [0]{\href }%
\providecommand \doibase [0]{https://doi.org/}%
\providecommand \selectlanguage [0]{\@gobble}%
\providecommand \bibinfo  [0]{\@secondoftwo}%
\providecommand \bibfield  [0]{\@secondoftwo}%
\providecommand \translation [1]{[#1]}%
\providecommand \BibitemOpen [0]{}%
\providecommand \bibitemStop [0]{}%
\providecommand \bibitemNoStop [0]{.\EOS\space}%
\providecommand \EOS [0]{\spacefactor3000\relax}%
\providecommand \BibitemShut  [1]{\csname bibitem#1\endcsname}%
\let\auto@bib@innerbib\@empty
\bibitem [{\citenamefont {Riess}\ \emph {et~al.}(1998)\citenamefont {Riess} \emph {et~al.}}]{Riess1998}%
  \BibitemOpen
  \bibfield  {author} {\bibinfo {author} {\bibfnamefont {A.~G.}\ \bibnamefont {Riess}} \emph {et~al.},\ }\href {https://doi.org/10.1086/300499} {\bibfield  {journal} {\bibinfo  {journal} {Astronomical Journal}\ }\textbf {\bibinfo {volume} {116}},\ \bibinfo {pages} {1009} (\bibinfo {year} {1998})},\ \Eprint {https://arxiv.org/abs/astro-ph/9805201} {astro-ph/9805201} \BibitemShut {NoStop}%
\bibitem [{\citenamefont {Perlmutter}\ \emph {et~al.}(1998)\citenamefont {Perlmutter}, \citenamefont {Aldering}, \citenamefont {Goldhaber}, \citenamefont {Knop}, \citenamefont {Nugent}, \citenamefont {Castro}, \citenamefont {Deustua}, \citenamefont {Fabbro}, \citenamefont {Goobar}, \citenamefont {Groom}, \citenamefont {Hook}, \citenamefont {Kim}, \citenamefont {Kim}, \citenamefont {Lee}, \citenamefont {Nunes}, \citenamefont {Pain}, \citenamefont {Pennypacker}, \citenamefont {Quimby}, \citenamefont {Lidman}, \citenamefont {Ellis}, \citenamefont {Irwin}, \citenamefont {McMahon}, \citenamefont {Ruiz-Lapuente}, \citenamefont {Walton}, \citenamefont {Schaefer}, \citenamefont {Boyle}, \citenamefont {Filippenko}, \citenamefont {Matheson}, \citenamefont {Fruchter}, \citenamefont {Panagia}, \citenamefont {Newberg},\ and\ \citenamefont {Couch}}]{Perlmutter1998}%
  \BibitemOpen
  \bibfield  {author} {\bibinfo {author} {\bibfnamefont {S.}~\bibnamefont {Perlmutter}}, \bibinfo {author} {\bibfnamefont {G.}~\bibnamefont {Aldering}}, \bibinfo {author} {\bibfnamefont {G.}~\bibnamefont {Goldhaber}}, \bibinfo {author} {\bibfnamefont {R.~A.}\ \bibnamefont {Knop}}, \bibinfo {author} {\bibfnamefont {P.}~\bibnamefont {Nugent}}, \bibinfo {author} {\bibfnamefont {P.~G.}\ \bibnamefont {Castro}}, \bibinfo {author} {\bibfnamefont {S.}~\bibnamefont {Deustua}}, \bibinfo {author} {\bibfnamefont {S.}~\bibnamefont {Fabbro}}, \bibinfo {author} {\bibfnamefont {A.}~\bibnamefont {Goobar}}, \bibinfo {author} {\bibfnamefont {D.~E.}\ \bibnamefont {Groom}}, \bibinfo {author} {\bibfnamefont {I.~M.}\ \bibnamefont {Hook}}, \bibinfo {author} {\bibfnamefont {A.~G.}\ \bibnamefont {Kim}}, \bibinfo {author} {\bibfnamefont {M.~Y.}\ \bibnamefont {Kim}}, \bibinfo {author} {\bibfnamefont {J.~C.}\ \bibnamefont {Lee}}, \bibinfo {author} {\bibfnamefont {N.~J.}\ \bibnamefont {Nunes}}, \bibinfo {author} {\bibfnamefont
  {R.}~\bibnamefont {Pain}}, \bibinfo {author} {\bibfnamefont {C.~R.}\ \bibnamefont {Pennypacker}}, \bibinfo {author} {\bibfnamefont {R.}~\bibnamefont {Quimby}}, \bibinfo {author} {\bibfnamefont {C.}~\bibnamefont {Lidman}}, \bibinfo {author} {\bibfnamefont {R.~S.}\ \bibnamefont {Ellis}}, \bibinfo {author} {\bibfnamefont {M.}~\bibnamefont {Irwin}}, \bibinfo {author} {\bibfnamefont {R.~G.}\ \bibnamefont {McMahon}}, \bibinfo {author} {\bibfnamefont {P.}~\bibnamefont {Ruiz-Lapuente}}, \bibinfo {author} {\bibfnamefont {N.}~\bibnamefont {Walton}}, \bibinfo {author} {\bibfnamefont {B.}~\bibnamefont {Schaefer}}, \bibinfo {author} {\bibfnamefont {B.~J.}\ \bibnamefont {Boyle}}, \bibinfo {author} {\bibfnamefont {A.~V.}\ \bibnamefont {Filippenko}}, \bibinfo {author} {\bibfnamefont {T.}~\bibnamefont {Matheson}}, \bibinfo {author} {\bibfnamefont {A.~S.}\ \bibnamefont {Fruchter}}, \bibinfo {author} {\bibfnamefont {N.}~\bibnamefont {Panagia}}, \bibinfo {author} {\bibfnamefont {H.~J.~M.}\ \bibnamefont {Newberg}},\ and\
  \bibinfo {author} {\bibfnamefont {W.~J.}\ \bibnamefont {Couch}},\ }\href {https://doi.org/10.1086/307221} {\bibfield  {journal} {\bibinfo  {journal} {Astrophys.J.517:565-586,1999}\ }\textbf {\bibinfo {volume} {517}},\ \bibinfo {pages} {565} (\bibinfo {year} {1998})},\ \Eprint {https://arxiv.org/abs/astro-ph/9812133} {arXiv:astro-ph/9812133 [astro-ph]} \BibitemShut {NoStop}%
\bibitem [{\citenamefont {Weinberg}(1989)}]{Weinberg1989}%
  \BibitemOpen
  \bibfield  {author} {\bibinfo {author} {\bibfnamefont {S.}~\bibnamefont {Weinberg}},\ }\href {https://doi.org/10.1103/RevModPhys.61.1"} {\bibfield  {journal} {\bibinfo  {journal} {Rev. Mod. Phys}\ ,\ \bibinfo {pages} {1}} (\bibinfo {year} {1989})}\BibitemShut {NoStop}%
\bibitem [{\citenamefont {Carroll}(2001)}]{Carroll2001}%
  \BibitemOpen
  \bibfield  {author} {\bibinfo {author} {\bibfnamefont {S.}~\bibnamefont {Carroll}},\ }\bibfield  {journal} {\bibinfo  {journal} {Living Reviews in Relativity}\ }\textbf {\bibinfo {volume} {4}},\ \href {https://doi.org/10.12942/lrr-2001-1} {10.12942/lrr-2001-1} (\bibinfo {year} {2001})\BibitemShut {NoStop}%
\bibitem [{\citenamefont {NOJIRI}\ and\ \citenamefont {ODINTSOV}(2007)}]{NOJIRI2007}%
  \BibitemOpen
  \bibfield  {author} {\bibinfo {author} {\bibfnamefont {S.}~\bibnamefont {NOJIRI}}\ and\ \bibinfo {author} {\bibfnamefont {S.~D.}\ \bibnamefont {ODINTSOV}},\ }\href {https://doi.org/10.1142/s0219887807001928} {\bibfield  {journal} {\bibinfo  {journal} {International Journal of Geometric Methods in Modern Physics}\ }\textbf {\bibinfo {volume} {04}},\ \bibinfo {pages} {115} (\bibinfo {year} {2007})}\BibitemShut {NoStop}%
\bibitem [{\citenamefont {Nojiri}\ and\ \citenamefont {Odintsov}(2003)}]{Nojiri2003}%
  \BibitemOpen
  \bibfield  {author} {\bibinfo {author} {\bibfnamefont {S.}~\bibnamefont {Nojiri}}\ and\ \bibinfo {author} {\bibfnamefont {S.~D.}\ \bibnamefont {Odintsov}},\ }\href {https://doi.org/10.1103/physrevd.68.123512} {\bibfield  {journal} {\bibinfo  {journal} {Physical Review D}\ }\textbf {\bibinfo {volume} {68}},\ \bibinfo {pages} {123512} (\bibinfo {year} {2003})}\BibitemShut {NoStop}%
\bibitem [{\citenamefont {Capozziello}\ and\ \citenamefont {De~Laurentis}(2011)}]{Capozziello2011}%
  \BibitemOpen
  \bibfield  {author} {\bibinfo {author} {\bibfnamefont {S.}~\bibnamefont {Capozziello}}\ and\ \bibinfo {author} {\bibfnamefont {M.}~\bibnamefont {De~Laurentis}},\ }\href {https://doi.org/10.1016/j.physrep.2011.09.003} {\bibfield  {journal} {\bibinfo  {journal} {Physics Reports}\ }\textbf {\bibinfo {volume} {509}},\ \bibinfo {pages} {167} (\bibinfo {year} {2011})}\BibitemShut {NoStop}%
\bibitem [{\citenamefont {Hu}\ and\ \citenamefont {Sawicki}(2007)}]{Hu2007}%
  \BibitemOpen
  \bibfield  {author} {\bibinfo {author} {\bibfnamefont {W.}~\bibnamefont {Hu}}\ and\ \bibinfo {author} {\bibfnamefont {I.}~\bibnamefont {Sawicki}},\ }\href {https://doi.org/10.1103/physrevd.76.064004} {\bibfield  {journal} {\bibinfo  {journal} {Physical Review D}\ }\textbf {\bibinfo {volume} {76}},\ \bibinfo {pages} {064004} (\bibinfo {year} {2007})}\BibitemShut {NoStop}%
\bibitem [{\citenamefont {Starobinsky}(2007)}]{Starobinsky2007}%
  \BibitemOpen
  \bibfield  {author} {\bibinfo {author} {\bibfnamefont {A.~A.}\ \bibnamefont {Starobinsky}},\ }\href {https://doi.org/10.1134/s0021364007150027} {\bibfield  {journal} {\bibinfo  {journal} {JETP Letters}\ }\textbf {\bibinfo {volume} {86}},\ \bibinfo {pages} {157} (\bibinfo {year} {2007})}\BibitemShut {NoStop}%
\bibitem [{\citenamefont {Appleby}\ and\ \citenamefont {Battye}(2007)}]{Appleby2007}%
  \BibitemOpen
  \bibfield  {author} {\bibinfo {author} {\bibfnamefont {S.~A.}\ \bibnamefont {Appleby}}\ and\ \bibinfo {author} {\bibfnamefont {R.~A.}\ \bibnamefont {Battye}},\ }\href {https://doi.org/10.1016/j.physletb.2007.08.037} {\bibfield  {journal} {\bibinfo  {journal} {Physics Letters B}\ }\textbf {\bibinfo {volume} {654}},\ \bibinfo {pages} {7} (\bibinfo {year} {2007})}\BibitemShut {NoStop}%
\bibitem [{\citenamefont {Caldwell}\ \emph {et~al.}()\citenamefont {Caldwell}, \citenamefont {Dave},\ and\ \citenamefont {Steinhardt}}]{Caldwell}%
  \BibitemOpen
  \bibfield  {author} {\bibinfo {author} {\bibfnamefont {R.~R.}\ \bibnamefont {Caldwell}}, \bibinfo {author} {\bibfnamefont {R.}~\bibnamefont {Dave}},\ and\ \bibinfo {author} {\bibfnamefont {P.~J.}\ \bibnamefont {Steinhardt}},\ }\href {https://doi.org/10.1103/physrevlett.80.1582} {\bibfield  {journal} {\bibinfo  {journal} {Physical Review Letters}\ }\textbf {\bibinfo {volume} {80}},\ \bibinfo {pages} {1582}}\BibitemShut {NoStop}%
\bibitem [{\citenamefont {Caldwell}(2002)}]{Caldwell2002}%
  \BibitemOpen
  \bibfield  {author} {\bibinfo {author} {\bibfnamefont {R.}~\bibnamefont {Caldwell}},\ }\href {https://doi.org/10.1016/s0370-2693(02)02589-3} {\bibfield  {journal} {\bibinfo  {journal} {Physics Letters B}\ }\textbf {\bibinfo {volume} {545}},\ \bibinfo {pages} {23} (\bibinfo {year} {2002})}\BibitemShut {NoStop}%
\bibitem [{\citenamefont {Brans}\ and\ \citenamefont {Dicke}(1961)}]{Brans1961}%
  \BibitemOpen
  \bibfield  {author} {\bibinfo {author} {\bibfnamefont {C.}~\bibnamefont {Brans}}\ and\ \bibinfo {author} {\bibfnamefont {R.~H.}\ \bibnamefont {Dicke}},\ }\href {https://doi.org/10.1103/physrev.124.925} {\bibfield  {journal} {\bibinfo  {journal} {Physical Review}\ }\textbf {\bibinfo {volume} {124}},\ \bibinfo {pages} {925} (\bibinfo {year} {1961})}\BibitemShut {NoStop}%
\bibitem [{\citenamefont {Hooft}(1993)}]{Hooft1993}%
  \BibitemOpen
  \bibfield  {author} {\bibinfo {author} {\bibfnamefont {G.~t.}\ \bibnamefont {Hooft}},\ }\href {https://doi.org/10.48550/ARXIV.GR-QC/9310026} {\bibinfo {title} {Dimensional reduction in quantum gravity}} (\bibinfo {year} {1993})\BibitemShut {NoStop}%
\bibitem [{\citenamefont {Susskind}(1995)}]{Susskind1995}%
  \BibitemOpen
  \bibfield  {author} {\bibinfo {author} {\bibfnamefont {L.}~\bibnamefont {Susskind}},\ }\href {https://doi.org/10.1063/1.531249} {\bibfield  {journal} {\bibinfo  {journal} {Journal of Mathematical Physics}\ }\textbf {\bibinfo {volume} {36}},\ \bibinfo {pages} {6377} (\bibinfo {year} {1995})}\BibitemShut {NoStop}%
\bibitem [{\citenamefont {Li}(2004)}]{Li2004}%
  \BibitemOpen
  \bibfield  {author} {\bibinfo {author} {\bibfnamefont {M.}~\bibnamefont {Li}},\ }\href {https://doi.org/10.1016/j.physletb.2004.10.014} {\bibfield  {journal} {\bibinfo  {journal} {Physics Letters B}\ }\textbf {\bibinfo {volume} {603}},\ \bibinfo {pages} {1} (\bibinfo {year} {2004})}\BibitemShut {NoStop}%
\bibitem [{\citenamefont {Wang}\ \emph {et~al.}(2017)\citenamefont {Wang}, \citenamefont {Wang},\ and\ \citenamefont {Li}}]{Wang2017}%
  \BibitemOpen
  \bibfield  {author} {\bibinfo {author} {\bibfnamefont {S.}~\bibnamefont {Wang}}, \bibinfo {author} {\bibfnamefont {Y.}~\bibnamefont {Wang}},\ and\ \bibinfo {author} {\bibfnamefont {M.}~\bibnamefont {Li}},\ }\href {https://doi.org/10.1016/j.physrep.2017.06.003} {\bibfield  {journal} {\bibinfo  {journal} {Physics Reports}\ }\textbf {\bibinfo {volume} {696}},\ \bibinfo {pages} {1} (\bibinfo {year} {2017})}\BibitemShut {NoStop}%
\bibitem [{\citenamefont {Solà}(2008)}]{Sola2008}%
  \BibitemOpen
  \bibfield  {author} {\bibinfo {author} {\bibfnamefont {J.}~\bibnamefont {Solà}},\ }\href {https://doi.org/10.1088/1751-8113/41/16/164066} {\bibfield  {journal} {\bibinfo  {journal} {Journal of Physics A: Mathematical and Theoretical}\ }\textbf {\bibinfo {volume} {41}},\ \bibinfo {pages} {164066} (\bibinfo {year} {2008})}\BibitemShut {NoStop}%
\bibitem [{\citenamefont {Shapiro}\ and\ \citenamefont {Solà}(2009)}]{Shapiro2009}%
  \BibitemOpen
  \bibfield  {author} {\bibinfo {author} {\bibfnamefont {I.~L.}\ \bibnamefont {Shapiro}}\ and\ \bibinfo {author} {\bibfnamefont {J.}~\bibnamefont {Solà}},\ }\href {https://doi.org/10.1016/j.physletb.2009.10.073} {\bibfield  {journal} {\bibinfo  {journal} {Physics Letters B}\ }\textbf {\bibinfo {volume} {682}},\ \bibinfo {pages} {105} (\bibinfo {year} {2009})}\BibitemShut {NoStop}%
\bibitem [{\citenamefont {Solà}(2011)}]{Sola2011}%
  \BibitemOpen
  \bibfield  {author} {\bibinfo {author} {\bibfnamefont {J.}~\bibnamefont {Solà}},\ }\href {https://doi.org/10.1088/1742-6596/283/1/012033} {\bibfield  {journal} {\bibinfo  {journal} {Journal of Physics: Conference Series}\ }\textbf {\bibinfo {volume} {283}},\ \bibinfo {pages} {012033} (\bibinfo {year} {2011})}\BibitemShut {NoStop}%
\bibitem [{\citenamefont {Easson}\ \emph {et~al.}(2011)\citenamefont {Easson}, \citenamefont {Frampton},\ and\ \citenamefont {Smoot}}]{Easson2011}%
  \BibitemOpen
  \bibfield  {author} {\bibinfo {author} {\bibfnamefont {D.~A.}\ \bibnamefont {Easson}}, \bibinfo {author} {\bibfnamefont {P.~H.}\ \bibnamefont {Frampton}},\ and\ \bibinfo {author} {\bibfnamefont {G.~F.}\ \bibnamefont {Smoot}},\ }\href {https://doi.org/10.1016/j.physletb.2010.12.025} {\bibfield  {journal} {\bibinfo  {journal} {Physics Letters B}\ }\textbf {\bibinfo {volume} {696}},\ \bibinfo {pages} {273} (\bibinfo {year} {2011})}\BibitemShut {NoStop}%
\bibitem [{\citenamefont {EASSON}\ \emph {et~al.}(2012)\citenamefont {EASSON}, \citenamefont {FRAMPTON},\ and\ \citenamefont {SMOOT}}]{EASSON2012}%
  \BibitemOpen
  \bibfield  {author} {\bibinfo {author} {\bibfnamefont {D.~A.}\ \bibnamefont {EASSON}}, \bibinfo {author} {\bibfnamefont {P.~H.}\ \bibnamefont {FRAMPTON}},\ and\ \bibinfo {author} {\bibfnamefont {G.~F.}\ \bibnamefont {SMOOT}},\ }\href {https://doi.org/10.1142/s0217751x12500662} {\bibfield  {journal} {\bibinfo  {journal} {International Journal of Modern Physics A}\ }\textbf {\bibinfo {volume} {27}},\ \bibinfo {pages} {1250066} (\bibinfo {year} {2012})}\BibitemShut {NoStop}%
\bibitem [{\citenamefont {Gohar}\ and\ \citenamefont {Salzano}(2023)}]{Gohar2023}%
  \BibitemOpen
  \bibfield  {author} {\bibinfo {author} {\bibfnamefont {H.}~\bibnamefont {Gohar}}\ and\ \bibinfo {author} {\bibfnamefont {V.}~\bibnamefont {Salzano}},\ }\href {https://doi.org/10.48550/ARXIV.2307.01768} {\bibinfo {title} {On the foundations of entropic cosmologies: inconsistencies, possible solutions and dead end signs}} (\bibinfo {year} {2023})\BibitemShut {NoStop}%
\bibitem [{\citenamefont {Tsallis}(1988)}]{Tsallis1988}%
  \BibitemOpen
  \bibfield  {author} {\bibinfo {author} {\bibfnamefont {C.}~\bibnamefont {Tsallis}},\ }\href {https://doi.org/10.1007/bf01016429} {\bibfield  {journal} {\bibinfo  {journal} {Journal of Statistical Physics}\ }\textbf {\bibinfo {volume} {52}},\ \bibinfo {pages} {479} (\bibinfo {year} {1988})}\BibitemShut {NoStop}%
\bibitem [{\citenamefont {Koivisto}\ \emph {et~al.}(2011)\citenamefont {Koivisto}, \citenamefont {Mota},\ and\ \citenamefont {Zumalacárregui}}]{Koivisto2011}%
  \BibitemOpen
  \bibfield  {author} {\bibinfo {author} {\bibfnamefont {T.~S.}\ \bibnamefont {Koivisto}}, \bibinfo {author} {\bibfnamefont {D.~F.}\ \bibnamefont {Mota}},\ and\ \bibinfo {author} {\bibfnamefont {M.}~\bibnamefont {Zumalacárregui}},\ }\href {https://doi.org/10.1088/1475-7516/2011/02/027} {\bibfield  {journal} {\bibinfo  {journal} {Journal of Cosmology and Astroparticle Physics}\ }\textbf {\bibinfo {volume} {2011}}\bibinfo  {number} { (02)},\ \bibinfo {pages} {027}}\BibitemShut {NoStop}%
\bibitem [{\citenamefont {Basilakos}\ \emph {et~al.}(2012)\citenamefont {Basilakos}, \citenamefont {Polarski},\ and\ \citenamefont {Solà}}]{Basilakos2012}%
  \BibitemOpen
\bibfield  {number} {  }\bibfield  {author} {\bibinfo {author} {\bibfnamefont {S.}~\bibnamefont {Basilakos}}, \bibinfo {author} {\bibfnamefont {D.}~\bibnamefont {Polarski}},\ and\ \bibinfo {author} {\bibfnamefont {J.}~\bibnamefont {Solà}},\ }\href {https://doi.org/10.1103/physrevd.86.043010} {\bibfield  {journal} {\bibinfo  {journal} {Physical Review D}\ }\textbf {\bibinfo {volume} {86}},\ \bibinfo {pages} {043010} (\bibinfo {year} {2012})}\BibitemShut {NoStop}%
\bibitem [{\citenamefont {Basilakos}\ and\ \citenamefont {Solà}(2014)}]{Basilakos2014}%
  \BibitemOpen
  \bibfield  {author} {\bibinfo {author} {\bibfnamefont {S.}~\bibnamefont {Basilakos}}\ and\ \bibinfo {author} {\bibfnamefont {J.}~\bibnamefont {Solà}},\ }\href {https://doi.org/10.1103/physrevd.90.023008} {\bibfield  {journal} {\bibinfo  {journal} {Physical Review D}\ }\textbf {\bibinfo {volume} {90}},\ \bibinfo {pages} {023008} (\bibinfo {year} {2014})}\BibitemShut {NoStop}%
\bibitem [{\citenamefont {Gohar}\ and\ \citenamefont {Salzano}(2024)}]{Gohar2024}%
  \BibitemOpen
  \bibfield  {author} {\bibinfo {author} {\bibfnamefont {H.}~\bibnamefont {Gohar}}\ and\ \bibinfo {author} {\bibfnamefont {V.}~\bibnamefont {Salzano}},\ }\href {https://doi.org/10.1103/physrevd.109.084075} {\bibfield  {journal} {\bibinfo  {journal} {Physical Review D}\ }\textbf {\bibinfo {volume} {109}},\ \bibinfo {pages} {084075} (\bibinfo {year} {2024})}\BibitemShut {NoStop}%
\bibitem [{\citenamefont {Komatsu}\ and\ \citenamefont {Kimura}(2014{\natexlab{a}})}]{Komatsu2014}%
  \BibitemOpen
  \bibfield  {author} {\bibinfo {author} {\bibfnamefont {N.}~\bibnamefont {Komatsu}}\ and\ \bibinfo {author} {\bibfnamefont {S.}~\bibnamefont {Kimura}},\ }\href {https://doi.org/10.1103/physrevd.89.123501} {\bibfield  {journal} {\bibinfo  {journal} {Physical Review D}\ }\textbf {\bibinfo {volume} {89}},\ \bibinfo {pages} {123501} (\bibinfo {year} {2014}{\natexlab{a}})}\BibitemShut {NoStop}%
\bibitem [{\citenamefont {Zhang}\ \emph {et~al.}(2014)\citenamefont {Zhang}, \citenamefont {Zhang}, \citenamefont {Yuan}, \citenamefont {Zhang},\ and\ \citenamefont {Sun}}]{Zhang:2012mp}%
  \BibitemOpen
  \bibfield  {author} {\bibinfo {author} {\bibfnamefont {C.}~\bibnamefont {Zhang}}, \bibinfo {author} {\bibfnamefont {H.}~\bibnamefont {Zhang}}, \bibinfo {author} {\bibfnamefont {S.}~\bibnamefont {Yuan}}, \bibinfo {author} {\bibfnamefont {T.-J.}\ \bibnamefont {Zhang}},\ and\ \bibinfo {author} {\bibfnamefont {Y.-C.}\ \bibnamefont {Sun}},\ }\href {https://doi.org/10.1088/1674-4527/14/10/002} {\bibfield  {journal} {\bibinfo  {journal} {Res. Astron. Astrophys.}\ }\textbf {\bibinfo {volume} {14}},\ \bibinfo {pages} {1221} (\bibinfo {year} {2014})},\ \Eprint {https://arxiv.org/abs/1207.4541} {arXiv:1207.4541 [astro-ph.CO]} \BibitemShut {NoStop}%
\bibitem [{\citenamefont {Simon}\ \emph {et~al.}(2005)\citenamefont {Simon}, \citenamefont {Verde},\ and\ \citenamefont {Jimenez}}]{Simon:2004tf}%
  \BibitemOpen
  \bibfield  {author} {\bibinfo {author} {\bibfnamefont {J.}~\bibnamefont {Simon}}, \bibinfo {author} {\bibfnamefont {L.}~\bibnamefont {Verde}},\ and\ \bibinfo {author} {\bibfnamefont {R.}~\bibnamefont {Jimenez}},\ }\href {https://doi.org/10.1103/PhysRevD.71.123001} {\bibfield  {journal} {\bibinfo  {journal} {Phys. Rev. D}\ }\textbf {\bibinfo {volume} {71}},\ \bibinfo {pages} {123001} (\bibinfo {year} {2005})},\ \Eprint {https://arxiv.org/abs/astro-ph/0412269} {arXiv:astro-ph/0412269} \BibitemShut {NoStop}%
\bibitem [{\citenamefont {Moresco}\ \emph {et~al.}(2012)\citenamefont {Moresco} \emph {et~al.}}]{Moresco:2012jh}%
  \BibitemOpen
  \bibfield  {author} {\bibinfo {author} {\bibfnamefont {M.}~\bibnamefont {Moresco}} \emph {et~al.},\ }\href {https://doi.org/10.1088/1475-7516/2012/08/006} {\bibfield  {journal} {\bibinfo  {journal} {JCAP}\ }\textbf {\bibinfo {volume} {08}},\ \bibinfo {pages} {006}},\ \Eprint {https://arxiv.org/abs/1201.3609} {arXiv:1201.3609 [astro-ph.CO]} \BibitemShut {NoStop}%
\bibitem [{\citenamefont {Alam}\ \emph {et~al.}(2017)\citenamefont {Alam} \emph {et~al.}}]{BOSS:2016wmc}%
  \BibitemOpen
  \bibfield  {author} {\bibinfo {author} {\bibfnamefont {S.}~\bibnamefont {Alam}} \emph {et~al.} (\bibinfo {collaboration} {BOSS}),\ }\href {https://doi.org/10.1093/mnras/stx721} {\bibfield  {journal} {\bibinfo  {journal} {Mon. Not. Roy. Astron. Soc.}\ }\textbf {\bibinfo {volume} {470}},\ \bibinfo {pages} {2617} (\bibinfo {year} {2017})},\ \Eprint {https://arxiv.org/abs/1607.03155} {arXiv:1607.03155 [astro-ph.CO]} \BibitemShut {NoStop}%
\bibitem [{\citenamefont {Moresco}\ \emph {et~al.}(2016)\citenamefont {Moresco}, \citenamefont {Pozzetti}, \citenamefont {Cimatti}, \citenamefont {Jimenez}, \citenamefont {Maraston}, \citenamefont {Verde}, \citenamefont {Thomas}, \citenamefont {Citro}, \citenamefont {Tojeiro},\ and\ \citenamefont {Wilkinson}}]{Moresco:2016mzx}%
  \BibitemOpen
  \bibfield  {author} {\bibinfo {author} {\bibfnamefont {M.}~\bibnamefont {Moresco}}, \bibinfo {author} {\bibfnamefont {L.}~\bibnamefont {Pozzetti}}, \bibinfo {author} {\bibfnamefont {A.}~\bibnamefont {Cimatti}}, \bibinfo {author} {\bibfnamefont {R.}~\bibnamefont {Jimenez}}, \bibinfo {author} {\bibfnamefont {C.}~\bibnamefont {Maraston}}, \bibinfo {author} {\bibfnamefont {L.}~\bibnamefont {Verde}}, \bibinfo {author} {\bibfnamefont {D.}~\bibnamefont {Thomas}}, \bibinfo {author} {\bibfnamefont {A.}~\bibnamefont {Citro}}, \bibinfo {author} {\bibfnamefont {R.}~\bibnamefont {Tojeiro}},\ and\ \bibinfo {author} {\bibfnamefont {D.}~\bibnamefont {Wilkinson}},\ }\href {https://doi.org/10.1088/1475-7516/2016/05/014} {\bibfield  {journal} {\bibinfo  {journal} {JCAP}\ }\textbf {\bibinfo {volume} {05}},\ \bibinfo {pages} {014}},\ \Eprint {https://arxiv.org/abs/1601.01701} {arXiv:1601.01701 [astro-ph.CO]} \BibitemShut {NoStop}%
\bibitem [{\citenamefont {Ratsimbazafy}\ \emph {et~al.}(2017)\citenamefont {Ratsimbazafy}, \citenamefont {Loubser}, \citenamefont {Crawford}, \citenamefont {Cress}, \citenamefont {Bassett}, \citenamefont {Nichol},\ and\ \citenamefont {V{\"a}is{\"a}nen}}]{Ratsimbazafy:2017vga}%
  \BibitemOpen
  \bibfield  {author} {\bibinfo {author} {\bibfnamefont {A.~L.}\ \bibnamefont {Ratsimbazafy}}, \bibinfo {author} {\bibfnamefont {S.~I.}\ \bibnamefont {Loubser}}, \bibinfo {author} {\bibfnamefont {S.~M.}\ \bibnamefont {Crawford}}, \bibinfo {author} {\bibfnamefont {C.~M.}\ \bibnamefont {Cress}}, \bibinfo {author} {\bibfnamefont {B.~A.}\ \bibnamefont {Bassett}}, \bibinfo {author} {\bibfnamefont {R.~C.}\ \bibnamefont {Nichol}},\ and\ \bibinfo {author} {\bibfnamefont {P.}~\bibnamefont {V{\"a}is{\"a}nen}},\ }\href {https://doi.org/10.1093/mnras/stx301} {\bibfield  {journal} {\bibinfo  {journal} {Mon. Not. Roy. Astron. Soc.}\ }\textbf {\bibinfo {volume} {467}},\ \bibinfo {pages} {3239} (\bibinfo {year} {2017})},\ \Eprint {https://arxiv.org/abs/1702.00418} {arXiv:1702.00418 [astro-ph.CO]} \BibitemShut {NoStop}%
\bibitem [{\citenamefont {Stern}\ \emph {et~al.}(2010)\citenamefont {Stern}, \citenamefont {Jimenez}, \citenamefont {Verde}, \citenamefont {Kamionkowski},\ and\ \citenamefont {Stanford}}]{Stern:2009ep}%
  \BibitemOpen
  \bibfield  {author} {\bibinfo {author} {\bibfnamefont {D.}~\bibnamefont {Stern}}, \bibinfo {author} {\bibfnamefont {R.}~\bibnamefont {Jimenez}}, \bibinfo {author} {\bibfnamefont {L.}~\bibnamefont {Verde}}, \bibinfo {author} {\bibfnamefont {M.}~\bibnamefont {Kamionkowski}},\ and\ \bibinfo {author} {\bibfnamefont {S.~A.}\ \bibnamefont {Stanford}},\ }\href {https://doi.org/10.1088/1475-7516/2010/02/008} {\bibfield  {journal} {\bibinfo  {journal} {JCAP}\ }\textbf {\bibinfo {volume} {02}},\ \bibinfo {pages} {008}},\ \Eprint {https://arxiv.org/abs/0907.3149} {arXiv:0907.3149 [astro-ph.CO]} \BibitemShut {NoStop}%
\bibitem [{\citenamefont {Jiao}\ \emph {et~al.}(2023)\citenamefont {Jiao}, \citenamefont {Borghi}, \citenamefont {Moresco},\ and\ \citenamefont {Zhang}}]{Jiao:2022aep}%
  \BibitemOpen
  \bibfield  {author} {\bibinfo {author} {\bibfnamefont {K.}~\bibnamefont {Jiao}}, \bibinfo {author} {\bibfnamefont {N.}~\bibnamefont {Borghi}}, \bibinfo {author} {\bibfnamefont {M.}~\bibnamefont {Moresco}},\ and\ \bibinfo {author} {\bibfnamefont {T.-J.}\ \bibnamefont {Zhang}},\ }\href {https://doi.org/10.3847/1538-4365/acbc77} {\bibfield  {journal} {\bibinfo  {journal} {Astrophys. J. Suppl.}\ }\textbf {\bibinfo {volume} {265}},\ \bibinfo {pages} {48} (\bibinfo {year} {2023})},\ \Eprint {https://arxiv.org/abs/2205.05701} {arXiv:2205.05701 [astro-ph.CO]} \BibitemShut {NoStop}%
\bibitem [{\citenamefont {Moresco}(2015)}]{Moresco:2015cya}%
  \BibitemOpen
  \bibfield  {author} {\bibinfo {author} {\bibfnamefont {M.}~\bibnamefont {Moresco}},\ }\href {https://doi.org/10.1093/mnrasl/slv037} {\bibfield  {journal} {\bibinfo  {journal} {Mon. Not. Roy. Astron. Soc.}\ }\textbf {\bibinfo {volume} {450}},\ \bibinfo {pages} {L16} (\bibinfo {year} {2015})},\ \Eprint {https://arxiv.org/abs/1503.01116} {arXiv:1503.01116 [astro-ph.CO]} \BibitemShut {NoStop}%
\bibitem [{\citenamefont {Lima}\ \emph {et~al.}(1997)\citenamefont {Lima}, \citenamefont {Zanchin},\ and\ \citenamefont {Brandenberger}}]{Lima1997}%
  \BibitemOpen
  \bibfield  {author} {\bibinfo {author} {\bibfnamefont {J.~A.~S.}\ \bibnamefont {Lima}}, \bibinfo {author} {\bibfnamefont {V.}~\bibnamefont {Zanchin}},\ and\ \bibinfo {author} {\bibfnamefont {R.}~\bibnamefont {Brandenberger}},\ }\href {https://doi.org/10.1093/mnras/291.1.l1} {\bibfield  {journal} {\bibinfo  {journal} {Monthly Notices of the Royal Astronomical Society}\ }\textbf {\bibinfo {volume} {291}},\ \bibinfo {pages} {L1} (\bibinfo {year} {1997})}\BibitemShut {NoStop}%
\bibitem [{\citenamefont {Jesus}\ \emph {et~al.}(2011)\citenamefont {Jesus}, \citenamefont {Oliveira}, \citenamefont {Basilakos},\ and\ \citenamefont {Lima}}]{Jesus2011}%
  \BibitemOpen
  \bibfield  {author} {\bibinfo {author} {\bibfnamefont {J.~F.}\ \bibnamefont {Jesus}}, \bibinfo {author} {\bibfnamefont {F.~A.}\ \bibnamefont {Oliveira}}, \bibinfo {author} {\bibfnamefont {S.}~\bibnamefont {Basilakos}},\ and\ \bibinfo {author} {\bibfnamefont {J.~A.~S.}\ \bibnamefont {Lima}},\ }\href {https://doi.org/10.1103/physrevd.84.063511} {\bibfield  {journal} {\bibinfo  {journal} {Physical Review D}\ }\textbf {\bibinfo {volume} {84}},\ \bibinfo {pages} {063511} (\bibinfo {year} {2011})}\BibitemShut {NoStop}%
\bibitem [{\citenamefont {Reis}(2003)}]{Reis2003}%
  \BibitemOpen
  \bibfield  {author} {\bibinfo {author} {\bibfnamefont {R.~R.~R.}\ \bibnamefont {Reis}},\ }\href {https://doi.org/10.1103/physrevd.67.087301} {\bibfield  {journal} {\bibinfo  {journal} {Physical Review D}\ }\textbf {\bibinfo {volume} {67}},\ \bibinfo {pages} {087301} (\bibinfo {year} {2003})}\BibitemShut {NoStop}%
\bibitem [{\citenamefont {Nayeri}\ and\ \citenamefont {Padmanabhan}(1998)}]{Nayeri1998}%
  \BibitemOpen
  \bibfield  {author} {\bibinfo {author} {\bibfnamefont {A.}~\bibnamefont {Nayeri}}\ and\ \bibinfo {author} {\bibfnamefont {T.}~\bibnamefont {Padmanabhan}},\ }\href {https://doi.org/10.48550/ARXIV.GR-QC/9807039} {\bibinfo {title} {A possible newtonian interpretation of relativistic cosmological perturbation theory}} (\bibinfo {year} {1998})\BibitemShut {NoStop}%
\bibitem [{\citenamefont {Komatsu}\ and\ \citenamefont {Kimura}(2014{\natexlab{b}})}]{Komatsu2014a}%
  \BibitemOpen
  \bibfield  {author} {\bibinfo {author} {\bibfnamefont {N.}~\bibnamefont {Komatsu}}\ and\ \bibinfo {author} {\bibfnamefont {S.}~\bibnamefont {Kimura}},\ }\href {https://doi.org/10.1103/physrevd.90.123516} {\bibfield  {journal} {\bibinfo  {journal} {Physical Review D}\ }\textbf {\bibinfo {volume} {90}},\ \bibinfo {pages} {123516} (\bibinfo {year} {2014}{\natexlab{b}})}\BibitemShut {NoStop}%
\bibitem [{\citenamefont {Solà Peracaula}\ \emph {et~al.}(2018)\citenamefont {Solà Peracaula}, \citenamefont {de Cruz Pérez},\ and\ \citenamefont {Gómez-Valent}}]{SolaPeracaula2018}%
  \BibitemOpen
  \bibfield  {author} {\bibinfo {author} {\bibfnamefont {J.}~\bibnamefont {Solà Peracaula}}, \bibinfo {author} {\bibfnamefont {J.}~\bibnamefont {de Cruz Pérez}},\ and\ \bibinfo {author} {\bibfnamefont {A.}~\bibnamefont {Gómez-Valent}},\ }\href {https://doi.org/10.1093/mnras/sty1253} {\bibfield  {journal} {\bibinfo  {journal} {Monthly Notices of the Royal Astronomical Society}\ }\textbf {\bibinfo {volume} {478}},\ \bibinfo {pages} {4357} (\bibinfo {year} {2018})}\BibitemShut {NoStop}%
\bibitem [{\citenamefont {Gómez-Valent}\ and\ \citenamefont {Solà}(2017)}]{GomezValent2017}%
  \BibitemOpen
  \bibfield  {author} {\bibinfo {author} {\bibfnamefont {A.}~\bibnamefont {Gómez-Valent}}\ and\ \bibinfo {author} {\bibfnamefont {J.}~\bibnamefont {Solà}},\ }\href {https://doi.org/10.1209/0295-5075/120/39001} {\bibfield  {journal} {\bibinfo  {journal} {EPL (Europhysics Letters)}\ }\textbf {\bibinfo {volume} {120}},\ \bibinfo {pages} {39001} (\bibinfo {year} {2017})}\BibitemShut {NoStop}%
\bibitem [{\citenamefont {Song}\ and\ \citenamefont {Percival}(2009)}]{Song2009}%
  \BibitemOpen
  \bibfield  {author} {\bibinfo {author} {\bibfnamefont {Y.-S.}\ \bibnamefont {Song}}\ and\ \bibinfo {author} {\bibfnamefont {W.~J.}\ \bibnamefont {Percival}},\ }\href {https://doi.org/10.1088/1475-7516/2009/10/004} {\bibfield  {journal} {\bibinfo  {journal} {Journal of Cosmology and Astroparticle Physics}\ }\textbf {\bibinfo {volume} {2009}}\bibinfo  {number} { (10)},\ \bibinfo {pages} {004}}\BibitemShut {NoStop}%
\bibitem [{\citenamefont {Achitouv}\ \emph {et~al.}(2017)\citenamefont {Achitouv}, \citenamefont {Blake}, \citenamefont {Carter}, \citenamefont {Koda},\ and\ \citenamefont {Beutler}}]{Achitouv2017}%
  \BibitemOpen
\bibfield  {number} {  }\bibfield  {author} {\bibinfo {author} {\bibfnamefont {I.}~\bibnamefont {Achitouv}}, \bibinfo {author} {\bibfnamefont {C.}~\bibnamefont {Blake}}, \bibinfo {author} {\bibfnamefont {P.}~\bibnamefont {Carter}}, \bibinfo {author} {\bibfnamefont {J.}~\bibnamefont {Koda}},\ and\ \bibinfo {author} {\bibfnamefont {F.}~\bibnamefont {Beutler}},\ }\href {https://doi.org/10.1103/physrevd.95.083502} {\bibfield  {journal} {\bibinfo  {journal} {Physical Review D}\ }\textbf {\bibinfo {volume} {95}},\ \bibinfo {pages} {083502} (\bibinfo {year} {2017})}\BibitemShut {NoStop}%
\bibitem [{\citenamefont {Okumura}\ \emph {et~al.}(2016)\citenamefont {Okumura}, \citenamefont {Hikage}, \citenamefont {Totani}, \citenamefont {Tonegawa}, \citenamefont {Okada}, \citenamefont {Glazebrook}, \citenamefont {Blake}, \citenamefont {Ferreira}, \citenamefont {More}, \citenamefont {Taruya}, \citenamefont {Tsujikawa}, \citenamefont {Akiyama}, \citenamefont {Dalton}, \citenamefont {Goto}, \citenamefont {Ishikawa}, \citenamefont {Iwamuro}, \citenamefont {Matsubara}, \citenamefont {Nishimichi}, \citenamefont {Ohta}, \citenamefont {Shimizu}, \citenamefont {Takahashi}, \citenamefont {Takato}, \citenamefont {Tamura}, \citenamefont {Yabe},\ and\ \citenamefont {Yoshida}}]{Okumura2016}%
  \BibitemOpen
  \bibfield  {author} {\bibinfo {author} {\bibfnamefont {T.}~\bibnamefont {Okumura}}, \bibinfo {author} {\bibfnamefont {C.}~\bibnamefont {Hikage}}, \bibinfo {author} {\bibfnamefont {T.}~\bibnamefont {Totani}}, \bibinfo {author} {\bibfnamefont {M.}~\bibnamefont {Tonegawa}}, \bibinfo {author} {\bibfnamefont {H.}~\bibnamefont {Okada}}, \bibinfo {author} {\bibfnamefont {K.}~\bibnamefont {Glazebrook}}, \bibinfo {author} {\bibfnamefont {C.}~\bibnamefont {Blake}}, \bibinfo {author} {\bibfnamefont {P.~G.}\ \bibnamefont {Ferreira}}, \bibinfo {author} {\bibfnamefont {S.}~\bibnamefont {More}}, \bibinfo {author} {\bibfnamefont {A.}~\bibnamefont {Taruya}}, \bibinfo {author} {\bibfnamefont {S.}~\bibnamefont {Tsujikawa}}, \bibinfo {author} {\bibfnamefont {M.}~\bibnamefont {Akiyama}}, \bibinfo {author} {\bibfnamefont {G.}~\bibnamefont {Dalton}}, \bibinfo {author} {\bibfnamefont {T.}~\bibnamefont {Goto}}, \bibinfo {author} {\bibfnamefont {T.}~\bibnamefont {Ishikawa}}, \bibinfo {author} {\bibfnamefont {F.}~\bibnamefont
  {Iwamuro}}, \bibinfo {author} {\bibfnamefont {T.}~\bibnamefont {Matsubara}}, \bibinfo {author} {\bibfnamefont {T.}~\bibnamefont {Nishimichi}}, \bibinfo {author} {\bibfnamefont {K.}~\bibnamefont {Ohta}}, \bibinfo {author} {\bibfnamefont {I.}~\bibnamefont {Shimizu}}, \bibinfo {author} {\bibfnamefont {R.}~\bibnamefont {Takahashi}}, \bibinfo {author} {\bibfnamefont {N.}~\bibnamefont {Takato}}, \bibinfo {author} {\bibfnamefont {N.}~\bibnamefont {Tamura}}, \bibinfo {author} {\bibfnamefont {K.}~\bibnamefont {Yabe}},\ and\ \bibinfo {author} {\bibfnamefont {N.}~\bibnamefont {Yoshida}},\ }\bibfield  {journal} {\bibinfo  {journal} {Publications of the Astronomical Society of Japan}\ }\textbf {\bibinfo {volume} {68}},\ \href {https://doi.org/10.1093/pasj/psw029} {10.1093/pasj/psw029} (\bibinfo {year} {2016})\BibitemShut {NoStop}%
\bibitem [{\citenamefont {Blake}\ \emph {et~al.}(2013)\citenamefont {Blake}, \citenamefont {Baldry}, \citenamefont {Bland-Hawthorn}, \citenamefont {Christodoulou}, \citenamefont {Colless}, \citenamefont {Conselice}, \citenamefont {Driver}, \citenamefont {Hopkins}, \citenamefont {Liske}, \citenamefont {Loveday}, \citenamefont {Norberg}, \citenamefont {Peacock}, \citenamefont {Poole},\ and\ \citenamefont {Robotham}}]{Blake2013}%
  \BibitemOpen
  \bibfield  {author} {\bibinfo {author} {\bibfnamefont {C.}~\bibnamefont {Blake}}, \bibinfo {author} {\bibfnamefont {I.~K.}\ \bibnamefont {Baldry}}, \bibinfo {author} {\bibfnamefont {J.}~\bibnamefont {Bland-Hawthorn}}, \bibinfo {author} {\bibfnamefont {L.}~\bibnamefont {Christodoulou}}, \bibinfo {author} {\bibfnamefont {M.}~\bibnamefont {Colless}}, \bibinfo {author} {\bibfnamefont {C.}~\bibnamefont {Conselice}}, \bibinfo {author} {\bibfnamefont {S.~P.}\ \bibnamefont {Driver}}, \bibinfo {author} {\bibfnamefont {A.~M.}\ \bibnamefont {Hopkins}}, \bibinfo {author} {\bibfnamefont {J.}~\bibnamefont {Liske}}, \bibinfo {author} {\bibfnamefont {J.}~\bibnamefont {Loveday}}, \bibinfo {author} {\bibfnamefont {P.}~\bibnamefont {Norberg}}, \bibinfo {author} {\bibfnamefont {J.~A.}\ \bibnamefont {Peacock}}, \bibinfo {author} {\bibfnamefont {G.~B.}\ \bibnamefont {Poole}},\ and\ \bibinfo {author} {\bibfnamefont {A.~S.~G.}\ \bibnamefont {Robotham}},\ }\href {https://doi.org/10.1093/mnras/stt1791} {\bibfield  {journal}
  {\bibinfo  {journal} {Monthly Notices of the Royal Astronomical Society}\ }\textbf {\bibinfo {volume} {436}},\ \bibinfo {pages} {3089} (\bibinfo {year} {2013})}\BibitemShut {NoStop}%
\bibitem [{\citenamefont {Marín}\ \emph {et~al.}(2015)\citenamefont {Marín}, \citenamefont {Beutler}, \citenamefont {Blake}, \citenamefont {Koda}, \citenamefont {Kazin},\ and\ \citenamefont {Schneider}}]{Marin2015}%
  \BibitemOpen
  \bibfield  {author} {\bibinfo {author} {\bibfnamefont {F.~A.}\ \bibnamefont {Marín}}, \bibinfo {author} {\bibfnamefont {F.}~\bibnamefont {Beutler}}, \bibinfo {author} {\bibfnamefont {C.}~\bibnamefont {Blake}}, \bibinfo {author} {\bibfnamefont {J.}~\bibnamefont {Koda}}, \bibinfo {author} {\bibfnamefont {E.}~\bibnamefont {Kazin}},\ and\ \bibinfo {author} {\bibfnamefont {D.~P.}\ \bibnamefont {Schneider}},\ }\href {https://doi.org/10.1093/mnras/stv2502} {\bibfield  {journal} {\bibinfo  {journal} {Monthly Notices of the Royal Astronomical Society}\ }\textbf {\bibinfo {volume} {455}},\ \bibinfo {pages} {4046} (\bibinfo {year} {2015})}\BibitemShut {NoStop}%
\bibitem [{\citenamefont {Bhattacharyya}\ and\ \citenamefont {Dasgupta}(2021)}]{Bhattacharyya2021}%
  \BibitemOpen
  \bibfield  {author} {\bibinfo {author} {\bibfnamefont {S.}~\bibnamefont {Bhattacharyya}}\ and\ \bibinfo {author} {\bibfnamefont {B.}~\bibnamefont {Dasgupta}},\ }\href {https://doi.org/10.1088/1475-7516/2021/07/023} {\bibfield  {journal} {\bibinfo  {journal} {Journal of Cosmology and Astroparticle Physics}\ }\textbf {\bibinfo {volume} {2021}}\bibinfo  {number} { (07)},\ \bibinfo {pages} {023}}\BibitemShut {NoStop}%
\bibitem [{\citenamefont {de~Mattia}\ \emph {et~al.}(2020)\citenamefont {de~Mattia}, \citenamefont {Ruhlmann-Kleider}, \citenamefont {Raichoor}, \citenamefont {Ross}, \citenamefont {Tamone}, \citenamefont {Zhao}, \citenamefont {Alam}, \citenamefont {Avila}, \citenamefont {Burtin}, \citenamefont {Bautista}, \citenamefont {Beutler}, \citenamefont {Brinkmann}, \citenamefont {Brownstein}, \citenamefont {Chapman}, \citenamefont {Chuang}, \citenamefont {Comparat}, \citenamefont {Bourboux}, \citenamefont {Dawson}, \citenamefont {de~la Macorra}, \citenamefont {Gil-Marín}, \citenamefont {Gonzalez-Perez}, \citenamefont {Gorgoni}, \citenamefont {Hou}, \citenamefont {Kong}, \citenamefont {Lin}, \citenamefont {Nadathur}, \citenamefont {Newman}, \citenamefont {Mueller}, \citenamefont {Percival}, \citenamefont {Rezaie}, \citenamefont {Rossi}, \citenamefont {Schneider}, \citenamefont {Tiwari}, \citenamefont {Vivek}, \citenamefont {Wang},\ and\ \citenamefont {Zhao}}]{Mattia2020}%
  \BibitemOpen
\bibfield  {number} {  }\bibfield  {author} {\bibinfo {author} {\bibfnamefont {A.}~\bibnamefont {de~Mattia}}, \bibinfo {author} {\bibfnamefont {V.}~\bibnamefont {Ruhlmann-Kleider}}, \bibinfo {author} {\bibfnamefont {A.}~\bibnamefont {Raichoor}}, \bibinfo {author} {\bibfnamefont {A.~J.}\ \bibnamefont {Ross}}, \bibinfo {author} {\bibfnamefont {A.}~\bibnamefont {Tamone}}, \bibinfo {author} {\bibfnamefont {C.}~\bibnamefont {Zhao}}, \bibinfo {author} {\bibfnamefont {S.}~\bibnamefont {Alam}}, \bibinfo {author} {\bibfnamefont {S.}~\bibnamefont {Avila}}, \bibinfo {author} {\bibfnamefont {E.}~\bibnamefont {Burtin}}, \bibinfo {author} {\bibfnamefont {J.}~\bibnamefont {Bautista}}, \bibinfo {author} {\bibfnamefont {F.}~\bibnamefont {Beutler}}, \bibinfo {author} {\bibfnamefont {J.}~\bibnamefont {Brinkmann}}, \bibinfo {author} {\bibfnamefont {J.~R.}\ \bibnamefont {Brownstein}}, \bibinfo {author} {\bibfnamefont {M.~J.}\ \bibnamefont {Chapman}}, \bibinfo {author} {\bibfnamefont {C.-H.}\ \bibnamefont {Chuang}}, \bibinfo
  {author} {\bibfnamefont {J.}~\bibnamefont {Comparat}}, \bibinfo {author} {\bibfnamefont {H.~d. M.~d.}\ \bibnamefont {Bourboux}}, \bibinfo {author} {\bibfnamefont {K.~S.}\ \bibnamefont {Dawson}}, \bibinfo {author} {\bibfnamefont {A.}~\bibnamefont {de~la Macorra}}, \bibinfo {author} {\bibfnamefont {H.}~\bibnamefont {Gil-Marín}}, \bibinfo {author} {\bibfnamefont {V.}~\bibnamefont {Gonzalez-Perez}}, \bibinfo {author} {\bibfnamefont {C.}~\bibnamefont {Gorgoni}}, \bibinfo {author} {\bibfnamefont {J.}~\bibnamefont {Hou}}, \bibinfo {author} {\bibfnamefont {H.}~\bibnamefont {Kong}}, \bibinfo {author} {\bibfnamefont {S.}~\bibnamefont {Lin}}, \bibinfo {author} {\bibfnamefont {S.}~\bibnamefont {Nadathur}}, \bibinfo {author} {\bibfnamefont {J.~A.}\ \bibnamefont {Newman}}, \bibinfo {author} {\bibfnamefont {E.-M.}\ \bibnamefont {Mueller}}, \bibinfo {author} {\bibfnamefont {W.~J.}\ \bibnamefont {Percival}}, \bibinfo {author} {\bibfnamefont {M.}~\bibnamefont {Rezaie}}, \bibinfo {author} {\bibfnamefont {G.}~\bibnamefont
  {Rossi}}, \bibinfo {author} {\bibfnamefont {D.~P.}\ \bibnamefont {Schneider}}, \bibinfo {author} {\bibfnamefont {P.}~\bibnamefont {Tiwari}}, \bibinfo {author} {\bibfnamefont {M.}~\bibnamefont {Vivek}}, \bibinfo {author} {\bibfnamefont {Y.}~\bibnamefont {Wang}},\ and\ \bibinfo {author} {\bibfnamefont {G.-B.}\ \bibnamefont {Zhao}},\ }\bibfield  {journal} {\bibinfo  {journal} {Monthly Notices of the Royal Astronomical Society}\ }\href {https://doi.org/10.1093/mnras/staa3891} {10.1093/mnras/staa3891} (\bibinfo {year} {2020})\BibitemShut {NoStop}%
\bibitem [{\citenamefont {Chapman}\ \emph {et~al.}(2022)\citenamefont {Chapman}, \citenamefont {Mohammad}, \citenamefont {Zhai}, \citenamefont {Percival}, \citenamefont {Tinker}, \citenamefont {Bautista}, \citenamefont {Brownstein}, \citenamefont {Burtin}, \citenamefont {Dawson}, \citenamefont {Gil-Marín}, \citenamefont {de la Macorra}, \citenamefont {Ross}, \citenamefont {Rossi}, \citenamefont {Schneider},\ and\ \citenamefont {Zhao}}]{Chapman2022}%
  \BibitemOpen
  \bibfield  {author} {\bibinfo {author} {\bibfnamefont {M.~J.}\ \bibnamefont {Chapman}}, \bibinfo {author} {\bibfnamefont {F.~G.}\ \bibnamefont {Mohammad}}, \bibinfo {author} {\bibfnamefont {Z.}~\bibnamefont {Zhai}}, \bibinfo {author} {\bibfnamefont {W.~J.}\ \bibnamefont {Percival}}, \bibinfo {author} {\bibfnamefont {J.~L.}\ \bibnamefont {Tinker}}, \bibinfo {author} {\bibfnamefont {J.~E.}\ \bibnamefont {Bautista}}, \bibinfo {author} {\bibfnamefont {J.~R.}\ \bibnamefont {Brownstein}}, \bibinfo {author} {\bibfnamefont {E.}~\bibnamefont {Burtin}}, \bibinfo {author} {\bibfnamefont {K.~S.}\ \bibnamefont {Dawson}}, \bibinfo {author} {\bibfnamefont {H.}~\bibnamefont {Gil-Marín}}, \bibinfo {author} {\bibfnamefont {A.}~\bibnamefont {de la Macorra}}, \bibinfo {author} {\bibfnamefont {A.~J.}\ \bibnamefont {Ross}}, \bibinfo {author} {\bibfnamefont {G.}~\bibnamefont {Rossi}}, \bibinfo {author} {\bibfnamefont {D.~P.}\ \bibnamefont {Schneider}},\ and\ \bibinfo {author} {\bibfnamefont {G.-B.}\ \bibnamefont {Zhao}},\
  }\href {https://doi.org/10.1093/mnras/stac1923} {\bibfield  {journal} {\bibinfo  {journal} {Monthly Notices of the Royal Astronomical Society}\ }\textbf {\bibinfo {volume} {516}},\ \bibinfo {pages} {617} (\bibinfo {year} {2022})}\BibitemShut {NoStop}%
\bibitem [{\citenamefont {Howlett}\ \emph {et~al.}(2015)\citenamefont {Howlett}, \citenamefont {Ross}, \citenamefont {Samushia}, \citenamefont {Percival},\ and\ \citenamefont {Manera}}]{Howlett2015}%
  \BibitemOpen
  \bibfield  {author} {\bibinfo {author} {\bibfnamefont {C.}~\bibnamefont {Howlett}}, \bibinfo {author} {\bibfnamefont {A.~J.}\ \bibnamefont {Ross}}, \bibinfo {author} {\bibfnamefont {L.}~\bibnamefont {Samushia}}, \bibinfo {author} {\bibfnamefont {W.~J.}\ \bibnamefont {Percival}},\ and\ \bibinfo {author} {\bibfnamefont {M.}~\bibnamefont {Manera}},\ }\href {https://doi.org/10.1093/mnras/stu2693} {\bibfield  {journal} {\bibinfo  {journal} {Monthly Notices of the Royal Astronomical Society}\ }\textbf {\bibinfo {volume} {449}},\ \bibinfo {pages} {848} (\bibinfo {year} {2015})}\BibitemShut {NoStop}%
\bibitem [{\citenamefont {Hawken}\ \emph {et~al.}(2017)\citenamefont {Hawken}, \citenamefont {Granett}, \citenamefont {Iovino}, \citenamefont {Guzzo}, \citenamefont {Peacock}, \citenamefont {de~la Torre}, \citenamefont {Garilli}, \citenamefont {Bolzonella}, \citenamefont {Scodeggio}, \citenamefont {Abbas}, \citenamefont {Adami}, \citenamefont {Bottini}, \citenamefont {Cappi}, \citenamefont {Cucciati}, \citenamefont {Davidzon}, \citenamefont {Fritz}, \citenamefont {Franzetti}, \citenamefont {Krywult}, \citenamefont {Le~Brun}, \citenamefont {Le~Fèvre}, \citenamefont {Maccagni}, \citenamefont {Małek}, \citenamefont {Marulli}, \citenamefont {Polletta}, \citenamefont {Pollo}, \citenamefont {Tasca}, \citenamefont {Tojeiro}, \citenamefont {Vergani}, \citenamefont {Zanichelli}, \citenamefont {Arnouts}, \citenamefont {Bel}, \citenamefont {Branchini}, \citenamefont {De~Lucia}, \citenamefont {Ilbert}, \citenamefont {Moscardini},\ and\ \citenamefont {Percival}}]{Hawken2017}%
  \BibitemOpen
  \bibfield  {author} {\bibinfo {author} {\bibfnamefont {A.~J.}\ \bibnamefont {Hawken}}, \bibinfo {author} {\bibfnamefont {B.~R.}\ \bibnamefont {Granett}}, \bibinfo {author} {\bibfnamefont {A.}~\bibnamefont {Iovino}}, \bibinfo {author} {\bibfnamefont {L.}~\bibnamefont {Guzzo}}, \bibinfo {author} {\bibfnamefont {J.~A.}\ \bibnamefont {Peacock}}, \bibinfo {author} {\bibfnamefont {S.}~\bibnamefont {de~la Torre}}, \bibinfo {author} {\bibfnamefont {B.}~\bibnamefont {Garilli}}, \bibinfo {author} {\bibfnamefont {M.}~\bibnamefont {Bolzonella}}, \bibinfo {author} {\bibfnamefont {M.}~\bibnamefont {Scodeggio}}, \bibinfo {author} {\bibfnamefont {U.}~\bibnamefont {Abbas}}, \bibinfo {author} {\bibfnamefont {C.}~\bibnamefont {Adami}}, \bibinfo {author} {\bibfnamefont {D.}~\bibnamefont {Bottini}}, \bibinfo {author} {\bibfnamefont {A.}~\bibnamefont {Cappi}}, \bibinfo {author} {\bibfnamefont {O.}~\bibnamefont {Cucciati}}, \bibinfo {author} {\bibfnamefont {I.}~\bibnamefont {Davidzon}}, \bibinfo {author} {\bibfnamefont
  {A.}~\bibnamefont {Fritz}}, \bibinfo {author} {\bibfnamefont {P.}~\bibnamefont {Franzetti}}, \bibinfo {author} {\bibfnamefont {J.}~\bibnamefont {Krywult}}, \bibinfo {author} {\bibfnamefont {V.}~\bibnamefont {Le~Brun}}, \bibinfo {author} {\bibfnamefont {O.}~\bibnamefont {Le~Fèvre}}, \bibinfo {author} {\bibfnamefont {D.}~\bibnamefont {Maccagni}}, \bibinfo {author} {\bibfnamefont {K.}~\bibnamefont {Małek}}, \bibinfo {author} {\bibfnamefont {F.}~\bibnamefont {Marulli}}, \bibinfo {author} {\bibfnamefont {M.}~\bibnamefont {Polletta}}, \bibinfo {author} {\bibfnamefont {A.}~\bibnamefont {Pollo}}, \bibinfo {author} {\bibfnamefont {L.~A.~M.}\ \bibnamefont {Tasca}}, \bibinfo {author} {\bibfnamefont {R.}~\bibnamefont {Tojeiro}}, \bibinfo {author} {\bibfnamefont {D.}~\bibnamefont {Vergani}}, \bibinfo {author} {\bibfnamefont {A.}~\bibnamefont {Zanichelli}}, \bibinfo {author} {\bibfnamefont {S.}~\bibnamefont {Arnouts}}, \bibinfo {author} {\bibfnamefont {J.}~\bibnamefont {Bel}}, \bibinfo {author} {\bibfnamefont
  {E.}~\bibnamefont {Branchini}}, \bibinfo {author} {\bibfnamefont {G.}~\bibnamefont {De~Lucia}}, \bibinfo {author} {\bibfnamefont {O.}~\bibnamefont {Ilbert}}, \bibinfo {author} {\bibfnamefont {L.}~\bibnamefont {Moscardini}},\ and\ \bibinfo {author} {\bibfnamefont {W.~J.}\ \bibnamefont {Percival}},\ }\href {https://doi.org/10.1051/0004-6361/201629678} {\bibfield  {journal} {\bibinfo  {journal} {Astronomy \&amp; Astrophysics}\ }\textbf {\bibinfo {volume} {607}},\ \bibinfo {pages} {A54} (\bibinfo {year} {2017})}\BibitemShut {NoStop}%
\bibitem [{\citenamefont {Mohammad}\ \emph {et~al.}(2018)\citenamefont {Mohammad}, \citenamefont {Bianchi}, \citenamefont {Percival}, \citenamefont {de~la Torre}, \citenamefont {Guzzo}, \citenamefont {Granett}, \citenamefont {Branchini}, \citenamefont {Bolzonella}, \citenamefont {Garilli}, \citenamefont {Scodeggio}, \citenamefont {Abbas}, \citenamefont {Adami}, \citenamefont {Bel}, \citenamefont {Bottini}, \citenamefont {Cappi}, \citenamefont {Cucciati}, \citenamefont {Davidzon}, \citenamefont {Franzetti}, \citenamefont {Fritz}, \citenamefont {Iovino}, \citenamefont {Krywult}, \citenamefont {Le~Brun}, \citenamefont {Le~Fèvre}, \citenamefont {Małek}, \citenamefont {Marulli}, \citenamefont {Polletta}, \citenamefont {Pollo}, \citenamefont {Tasca}, \citenamefont {Tojeiro}, \citenamefont {Vergani}, \citenamefont {Zanichelli}, \citenamefont {Arnouts}, \citenamefont {Coupon}, \citenamefont {De~Lucia}, \citenamefont {Ilbert}, \citenamefont {Moscardini},\ and\ \citenamefont {Moutard}}]{Mohammad2018}%
  \BibitemOpen
  \bibfield  {author} {\bibinfo {author} {\bibfnamefont {F.~G.}\ \bibnamefont {Mohammad}}, \bibinfo {author} {\bibfnamefont {D.}~\bibnamefont {Bianchi}}, \bibinfo {author} {\bibfnamefont {W.~J.}\ \bibnamefont {Percival}}, \bibinfo {author} {\bibfnamefont {S.}~\bibnamefont {de~la Torre}}, \bibinfo {author} {\bibfnamefont {L.}~\bibnamefont {Guzzo}}, \bibinfo {author} {\bibfnamefont {B.~R.}\ \bibnamefont {Granett}}, \bibinfo {author} {\bibfnamefont {E.}~\bibnamefont {Branchini}}, \bibinfo {author} {\bibfnamefont {M.}~\bibnamefont {Bolzonella}}, \bibinfo {author} {\bibfnamefont {B.}~\bibnamefont {Garilli}}, \bibinfo {author} {\bibfnamefont {M.}~\bibnamefont {Scodeggio}}, \bibinfo {author} {\bibfnamefont {U.}~\bibnamefont {Abbas}}, \bibinfo {author} {\bibfnamefont {C.}~\bibnamefont {Adami}}, \bibinfo {author} {\bibfnamefont {J.}~\bibnamefont {Bel}}, \bibinfo {author} {\bibfnamefont {D.}~\bibnamefont {Bottini}}, \bibinfo {author} {\bibfnamefont {A.}~\bibnamefont {Cappi}}, \bibinfo {author} {\bibfnamefont
  {O.}~\bibnamefont {Cucciati}}, \bibinfo {author} {\bibfnamefont {I.}~\bibnamefont {Davidzon}}, \bibinfo {author} {\bibfnamefont {P.}~\bibnamefont {Franzetti}}, \bibinfo {author} {\bibfnamefont {A.}~\bibnamefont {Fritz}}, \bibinfo {author} {\bibfnamefont {A.}~\bibnamefont {Iovino}}, \bibinfo {author} {\bibfnamefont {J.}~\bibnamefont {Krywult}}, \bibinfo {author} {\bibfnamefont {V.}~\bibnamefont {Le~Brun}}, \bibinfo {author} {\bibfnamefont {O.}~\bibnamefont {Le~Fèvre}}, \bibinfo {author} {\bibfnamefont {K.}~\bibnamefont {Małek}}, \bibinfo {author} {\bibfnamefont {F.}~\bibnamefont {Marulli}}, \bibinfo {author} {\bibfnamefont {M.}~\bibnamefont {Polletta}}, \bibinfo {author} {\bibfnamefont {A.}~\bibnamefont {Pollo}}, \bibinfo {author} {\bibfnamefont {L.~A.~M.}\ \bibnamefont {Tasca}}, \bibinfo {author} {\bibfnamefont {R.}~\bibnamefont {Tojeiro}}, \bibinfo {author} {\bibfnamefont {D.}~\bibnamefont {Vergani}}, \bibinfo {author} {\bibfnamefont {A.}~\bibnamefont {Zanichelli}}, \bibinfo {author} {\bibfnamefont
  {S.}~\bibnamefont {Arnouts}}, \bibinfo {author} {\bibfnamefont {J.}~\bibnamefont {Coupon}}, \bibinfo {author} {\bibfnamefont {G.}~\bibnamefont {De~Lucia}}, \bibinfo {author} {\bibfnamefont {O.}~\bibnamefont {Ilbert}}, \bibinfo {author} {\bibfnamefont {L.}~\bibnamefont {Moscardini}},\ and\ \bibinfo {author} {\bibfnamefont {T.}~\bibnamefont {Moutard}},\ }\href {https://doi.org/10.1051/0004-6361/201833853} {\bibfield  {journal} {\bibinfo  {journal} {Astronomy \&amp; Astrophysics}\ }\textbf {\bibinfo {volume} {619}},\ \bibinfo {pages} {A17} (\bibinfo {year} {2018})}\BibitemShut {NoStop}%
\bibitem [{\citenamefont {Jullo}\ \emph {et~al.}(2019)\citenamefont {Jullo}, \citenamefont {de~la Torre}, \citenamefont {Cousinou}, \citenamefont {Escoffier}, \citenamefont {Giocoli}, \citenamefont {Metcalf}, \citenamefont {Comparat}, \citenamefont {Shan}, \citenamefont {Makler}, \citenamefont {Kneib}, \citenamefont {Prada}, \citenamefont {Yepes},\ and\ \citenamefont {Gottlöber}}]{Jullo2019}%
  \BibitemOpen
  \bibfield  {author} {\bibinfo {author} {\bibfnamefont {E.}~\bibnamefont {Jullo}}, \bibinfo {author} {\bibfnamefont {S.}~\bibnamefont {de~la Torre}}, \bibinfo {author} {\bibfnamefont {M.-C.}\ \bibnamefont {Cousinou}}, \bibinfo {author} {\bibfnamefont {S.}~\bibnamefont {Escoffier}}, \bibinfo {author} {\bibfnamefont {C.}~\bibnamefont {Giocoli}}, \bibinfo {author} {\bibfnamefont {R.~B.}\ \bibnamefont {Metcalf}}, \bibinfo {author} {\bibfnamefont {J.}~\bibnamefont {Comparat}}, \bibinfo {author} {\bibfnamefont {H.-Y.}\ \bibnamefont {Shan}}, \bibinfo {author} {\bibfnamefont {M.}~\bibnamefont {Makler}}, \bibinfo {author} {\bibfnamefont {J.-P.}\ \bibnamefont {Kneib}}, \bibinfo {author} {\bibfnamefont {F.}~\bibnamefont {Prada}}, \bibinfo {author} {\bibfnamefont {G.}~\bibnamefont {Yepes}},\ and\ \bibinfo {author} {\bibfnamefont {S.}~\bibnamefont {Gottlöber}},\ }\href {https://doi.org/10.1051/0004-6361/201834629} {\bibfield  {journal} {\bibinfo  {journal} {Astronomy \&amp; Astrophysics}\ }\textbf {\bibinfo {volume}
  {627}},\ \bibinfo {pages} {A137} (\bibinfo {year} {2019})}\BibitemShut {NoStop}%
\end{thebibliography}%
\end{document}